\documentclass[letterpaper,twocolumn,10pt]{article}
\usepackage{usenix-2020-09}

\usepackage{tikz}
\usepackage{amsmath}

\usepackage{xurl}
\usepackage{textcomp}
\usepackage{setspace}
\usepackage{amsfonts}
\usepackage{makecell}
\usepackage{enumerate}
\usepackage[ruled]{algorithm2e}
\usepackage{algpseudocode}
\usepackage{graphics}
\usepackage{amsthm}

\usepackage{booktabs}
\usepackage{xparse} 
\usepackage{xspace}
\usepackage{multirow}
\usepackage{balance}
\usepackage{diagbox}
\usepackage{bbding}
\usepackage{utfsym}
\usepackage{colortbl}
\usepackage{booktabs}
\usepackage{mdframed}
\usepackage{tikz}
\usepackage{amsmath}
\usepackage{hyperref}
\usepackage{cleveref}
\hypersetup{
    colorlinks = true,
    linkcolor = blue,
    anchorcolor = blue,
    citecolor = blue,
    filecolor = blue,
    urlcolor = blue
}

\usepackage{tabularx,lipsum,environ,amssymb}

\makeatletter
\newcommand{\problemtitle}[1]{\gdef\@problemtitle{#1}}
\newcommand{\probleminput}[1]{\gdef\@probleminput{#1}}
\newcommand{\problemquestion}[1]{\gdef\@problemquestion{#1}}
\NewEnviron{problem}{
  \problemtitle{}\probleminput{}\problemquestion{}
  \BODY
  \par\addvspace{.5\baselineskip}
  \noindent
  \begin{tabularx}{\linewidth}{@{\hspace{\parindent}} l X c}
    \multicolumn{2}{@{\hspace{\parindent}}l}{\@problemtitle} \\
    \textbf{Input:} & \@probleminput \\
    \textbf{Question:} & \@problemquestion
  \end{tabularx}
  \par\addvspace{.5\baselineskip}
}
\makeatother

\newcommand{\xmark}{\ding{55}}%
\makeatother
\newcommand{\tickYes}{\ding{52}}
\newcommand{\tickNo}{\xmark}

\usepackage{pifont}

\renewcommand{\paragraph}[1]{\vspace{0.05in}\noindent{\bf{#1}.}}

\begin{document}
%
\title{The Cost of {Performance}: Breaking ThreadX with Kernel Object Masquerading Attacks}

\author{
 \rm Xinhui Shao$^\dagger$,\;
 Zhen Ling$^\dagger$\thanks{Corresponding author: Prof. Zhen Ling of Southeast University, China.},\;
 Yue Zhang$^\ddagger$,\;
 Huaiyu Yan$^\dagger$,\;
 Yumeng Wei$^\dagger$,\;
 Lan Luo$^\P$,\;
 Zixia Liu$^\P$,\;\\
 Junzhou Luo$^\dagger$,\;
 Xinwen Fu$^\S$\;
 \\
 $^\dagger$  Southeast University, Email: \{xinhuishao, zhenling, huaiyu\_yan, yumeng5, jluo\}@seu.edu.cn\\
 $^\ddagger$ Drexel University, Email: zyueinfosec@gmail.com\\
 $^\P$ Anhui University of Technology, Email: \{lluo, zxliu\}@ahut.edu.cn\\
 $^\S$ University of Massachusetts Lowell, Email: xinwen\_fu@uml.edu
}

\maketitle

\begin{abstract}
Microcontroller-based IoT devices often use embedded real-time operating systems (RTOSs).
Vulnerabilities in these embedded RTOSs can lead to compromises of those IoT devices. Despite the significance of security protections, the absence of standardized security guidelines results in various levels of security risk across RTOS implementations. Our initial analysis reveals that popular RTOSs such as FreeRTOS lack essential security protections.
While Zephyr OS and ThreadX are designed and implemented with essential security protections, our closer examination uncovers significant differences in their implementations of system call parameter sanitization. We identify a performance optimization practice in ThreadX that introduces security vulnerabilities, allowing for the circumvention of parameter sanitization processes. Leveraging this insight, we introduce a novel attack named the Kernel Object Masquerading (KOM) Attack (as the attacker needs to manipulate one or multiple kernel objects through carefully selected system calls to launch the attack), demonstrating how attackers can exploit these vulnerabilities to access sensitive fields within kernel objects, potentially leading to unauthorized data manipulation, privilege escalation, or system compromise. We introduce an automated approach involving under-constrained symbolic execution to identify the KOM attacks and to understand the implications. Experimental results demonstrate the feasibility of KOM attacks on ThreadX-powered platforms. We reported our findings to the vendors, who recognized the vulnerabilities, with Amazon and Microsoft acknowledging our contribution on their websites.
\looseness=-1
\end{abstract}

\section{Introduction}
\label{sec:Introduction}

Real-time operating systems (RTOSs), such as FreeRTOS and ThreadX \cite{DBLP:conf/mobilecloud/JainL23,lamie2019real,DBLP:conf/sa/BorgesPSGC21,kamboh2006demonstration,heydarzadeh2022precise}, are widely utilized in microcontroller (MCU) based Internet of Things (IoT) devices to enable efficient multitasking. 
These MCU-oriented RTOSs play a critical role in the rapidly expanding IoT ecosystem, supporting a range of applications including smart homes, industrial automation, automotive systems, and medical devices. For simplicity, unless explicitly stated otherwise, the term RTOS in this paper refers to the MCU oriented RTOS. 
\looseness=-1

Despite their pivotal role, the lack of standardized guidelines has resulted in diverse security implementations across MCU-oriented RTOSs, leading to varying levels of security risk. Among the RTOSs we investigated, a primary security objective is to isolate the kernel from user threads.
To evaluate how effectively these RTOSs achieve this, we focus on three key protections: privilege separation, memory access control and system call parameter sanitization. These protections, well-established in full-fledged operating systems like Linux, are essential for creating and maintaining a secure boundary between the kernel and user threads, involving both hardware and software protections.
Hardware protections, which include privilege separation and memory access control~\cite{freertos-mpu-protection,threadx-mpu-protection,zephyr-mpu-protection, zhouunderstanding}, enforce memory isolation between the kernel and user threads. Software protections, specifically system call parameter sanitization, ensure secure interactions between the kernel and user threads, enhancing existing memory isolation by defending against vulnerabilities in higher-privileged code that could be exploited by malicious inputs, such as confused deputy attacks~\cite{DBLP:conf/ndss/MachiryGSSSWBCK17,DBLP:conf/asplos/CheckowayS13,DBLP:conf/uss/SuciuMSS20}. Our initial security analysis reveals that many RTOSs (e.g., FreeRTOS) lack at least one of these three critical security protections, making them vulnerable to various attacks.
Among the RTOSs we examined, Zephyr OS and ThreadX are the only ones that implement all these essential security protections. \looseness=-1

While both ThreadX and Zephyr OS implement essential security protections, their implementations of system call sanitization differ significantly, particularly in the validation of kernel object pointers. Our experiments indicate that ThreadX achieves higher performance in system call sanitization compared to Zephyr OS, which employs a method similar to those used in full-fledged operating systems. However, this performance advantage raises concerns about the security effectiveness of ThreadX.
Upon examining their implementations in detail, we found that both Zephyr OS and ThreadX use address validation and semantic validation to sanitize kernel object pointers. Zephyr OS employs a fine-grained strategy that securely stores kernel object pointer addresses and their semantics (e.g., the type of the kernel object) in kernel memory and verifies whether the stored information matches the kernel object pointers during system call execution. In contrast, ThreadX adopts a more coarse-grained but efficient approach: (i) Instead of saving all addresses and exhaustively checking them against input kernel object pointers, it only verifies that the input kernel object pointer falls outside the isolated memory of the currently running thread (i.e., address validation); (ii) rather than storing all semantics and exhaustively checking all fields against the stored semantics, it checks only specific fields (referred to as \textit{condition fields}) of the pointed kernel objects to meet certain criteria, such as confirming that the kernel objects have the expected types and states (i.e., semantic validation).\looseness=-1

In this paper, we aim to investigate whether the performance optimization practices in ThreadX could introduce security vulnerabilities and provide an in-depth analysis of its security.
We make two observations with security implications:  
\textbf{(O-I)} address validation requires that the input kernel object pointer resides beyond the isolated memory scope of the presently executing thread. However, any kernel object pointers originally intended for kernel memory access, rather than the isolated memory of the running thread, fulfill this criterion.
\textbf{(O-II)} we found that certain ThreadX system calls permit threads to alter specific fields (referred to as \textit{modifiable fields}) of a kernel object. Consequently, a thread can modify these fields of a valid kernel object to particular values, masquerading as the condition fields of the kernel object used in another system call, thus meeting semantic validation.\looseness=-1

These two observations on performance optimizations in system call sanitization motivate us to develop a method aimed at circumventing the parameter sanitization process for kernel object pointers through sequentially invoking carefully chosen system calls: A user thread can set the modifiable fields of a kernel object (by invoking system call A) to values that meet the semantic requirements of system call B. If the modified fields align with the condition fields checked by system call B, the kernel object pointer passed to system call B can reference a “virtual” kernel object (referred to as a \textit{forged kernel object}) that masquerades as a valid kernel object. This forged kernel object is capable of bypassing the parameter sanitization process of system call B. First, the kernel object pointer referencing the forged kernel object can pass address validation (See \textbf{O-I}). Second, by possessing condition fields identical to those required by system call B, the forged kernel object can satisfy semantic validation (See \textbf{O-II}). \looseness=-1

Building on the circumvention technique of parameter sanitization, we propose a novel Kernel Object Masquerading (KOM) attack. This attack aims to gain unauthorized access to sensitive memory, typically with elevated privileges, by invoking a sequence of carefully selected system calls. This attack constructs a chain of forged kernel objects that masquerade as valid kernel objects,
enabling attackers to overwrite a pointer field in a target kernel object (referred to as the \textit{accomplice kernel object}). This pointer can then be used to achieve arbitrary memory access, potentially leading to various severe consequences, such as unauthorized data manipulation, privilege escalation, or even complete system compromise. Although manually identifying the vulnerable system calls is feasible, it is labor-intensive and prone to errors. To address this, we introduce an automated approach using an under-constrained symbolic execution engine~\cite{DBLP:conf/usenix/RamosE16, DBLP:journals/ase/DengLR12}. This engine symbolically executes each system call, analyzing memory operations and extracting program dependencies related to modifiable fields to efficiently identify vulnerable system calls.\looseness=-1

To further understand the implications of our attack, we conduct a comprehensive evaluation of the KOM attack.
Our experimental results reveal that more than half of the system calls can be exploited to conduct a KOM attack. Furthermore, we successfully executed KOM attacks on three distinct platforms powered by ThreadX, namely STM, OLIMEX, and Wilderness Labs. We promptly reported our findings, accompanied by a Proof of Concept, to Microsoft. Microsoft acknowledged our contribution on their websites.\looseness=-1

Our major contributions are summarized as follows.
\begin{itemize}
\item 
\textbf{Understanding Status of RTOSs:} We are the first to conduct a comprehensive security analysis of the protective mechanisms in mainstream RTOSs. 
Our findings have been corroborated by various vendors such as Microsoft and Amazon (the proprietor of FreeRTOS). 

\item \textbf{Attacks with Novel Insights:} We present a novel attack, the Kernel Object Masquerading (KOM) attack, which exploits a newly discovered vulnerability in ThreadX, a widely used RTOS known for its robust security features. KOM attacks pose significant security threats, such as privilege escalation and unauthorized data manipulation, to all devices (e.g., STM, OLIMEX, and Wilderness Labs) running ThreadX.

\item
\textbf{Novel Techniques with Practical Impacts:} 
We have developed an automated technique to identify vulnerable system calls in KOM attacks using under-constrained symbolic execution, focusing on memory operations involving kernel objects. This method is generalizable and can be applied to uncover vulnerabilities in other RTOSs.

\item \textbf{Empirical Study with New Findings:} 
Our experimental results reveal that nearly half of the system calls (31) can be utilized for KOM attacks. Furthermore, we discovered that KOM attacks can be effective across various attack environments when using carefully selected system calls. Our experimental results demonstrate the feasibility of KOM attacks on ThreadX-powered real-world devices. 
\end{itemize}

\section{Background}

\subsection{RTOS Overview}
\label{subsec::bg::rtos}

A Real-Time Operating System (RTOS) is a specialized operating system that provides a suite of kernel services designed to meet real-time requirements. In this paper, we specifically focus on the RTOS implementations tailored for low-end embedded systems equipped with resource-constrained MCUs. To gain a comprehensive bottom-up understanding of RTOS, we delve into both its hardware and software components. \looseness=-1



\paragraph{Hardware} From a hardware perspective, MCUs provide a low-power environment for RTOS-based applications, supporting essential system management features like privilege management and memory access control. The Cortex-M series is a prominent family of 32-bit processors specifically designed for use in MCUs within the ARM architecture. These processors support privilege separation with two levels of execution privilege: privileged and unprivileged. Privileged code has access to all system resources while unprivileged code has limited access. Unprivileged code can transfer control to privileged code by executing the \texttt{svc} instruction to make a supervisor call (i.e., system call)~\cite{armv7-m,armv8-m}. 

For memory access control, most Cortex-M series processors incorporate the Memory Protection Unit (MPU)~\cite{armv8-m-mpu}, a specialized security hardware component. The MPU is a lightweight alternative to the Memory Management Unit (MMU), designed for systems without virtual memory. Privileged code can configure the MPU control register at runtime to set specific memory access permissions (i.e., read, write, and execute) or to enable/disable the MPU.  \looseness=-1

\paragraph{Software} From a software perspective, an RTOS-based application consists of a kernel and one or more user threads (often referred to as tasks in RTOSs such as FreeRTOS). The kernel provides essential services to user threads, including multitasking and efficient resource management, through a system call or function call mechanism.
The kernel typically maintains a set of data structures (i.e., kernel objects) of various types, such as threads and timers, which are used to manage these services and are accessible exclusively by the kernel itself. User threads can request specific kernel services to create and manipulate these kernel objects. For example, in ThreadX \cite{DBLP:conf/sa/BorgesPSGC21}, a user thread can create a thread kernel object and subsequently adjust its scheduling time slice. The user thread requests the kernel to create a thread kernel object and then modify the \texttt{tx\_thread\_time\_slice} member variable within the corresponding kernel object. 
It is worth noting that all the RTOSs we investigated (as shown in \autoref{tab:protections-rtos}) expose kernel object pointers to user threads. This design allows user threads to efficiently access memory and locate their assigned kernel objects.\looseness=-1
\begin{figure}[ht]
\centering

\includegraphics[width=0.9\linewidth]{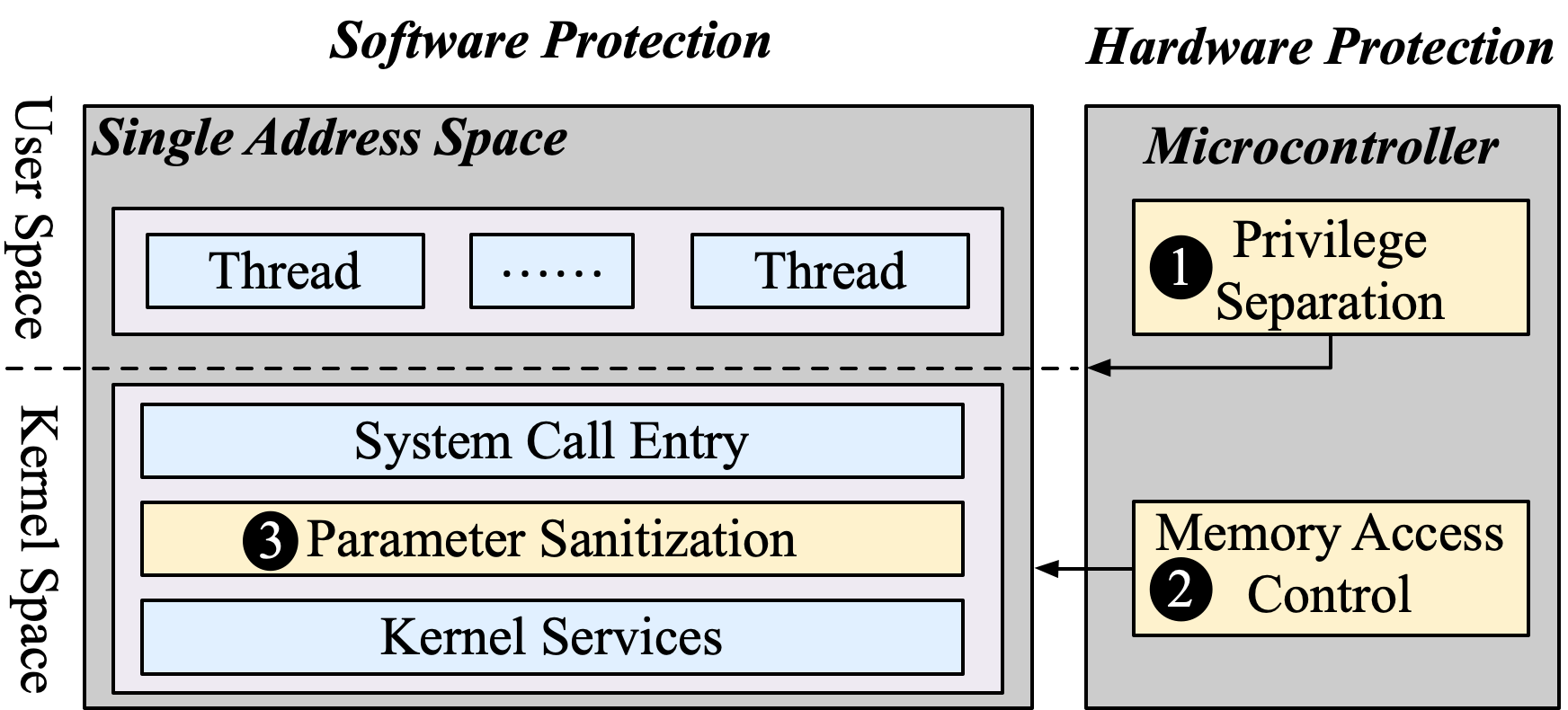}
\caption{Security Protections of RTOS}
\label{fig:protection-on-rtos}
\vspace{-5mm}
\end{figure}

\subsection{RTOS Security Protections}
\label{subsec::bg::protect}
In this paper, we focus on the three key security protections designed to safeguard the RTOS kernel from potentially compromised user threads. These protections also play a crucial role in full-fledged operating systems such as Linux and Windows.
\autoref{fig:protection-on-rtos} illustrates the protections employed by RTOSs, categorized into hardware protections (HP) and software protections (SP). 


\begin{itemize}
    \item [\ding{182}] \textbf{Privilege Separation (HP)} divides the software into privileged and unprivileged software. Typically, with privilege separation enabled, user threads execute at the unprivileged level while the RTOS kernel executes at the privileged level. This ensures that less-privileged threads are restricted from unauthorized accesses to sensitive system resources, such as the MPU's control registers.\looseness=-1 
    \item [\ding{183}] \textbf{Memory Access Control (HP)} enforces fine-grained memory access restrictions on developer-defined memory regions using the MPU. Generally, user threads are prevented from accessing kernel memory and the memory of other user threads~\cite{freertos-mpu-protection,threadx-mpu-protection,zephyr-mpu-protection}. This is achieved by reconfiguring the MPU when the kernel switches the running thread, allowing only the memory owned by the currently running thread to be accessible. In addition, with the MPU, RTOSs can also enforce further access restrictions for code and data memory, such as Data Execution Prevention (DEP)~\cite{DEP}, for both user threads and the RTOS kernel~\cite{zhouunderstanding}.\looseness=-1
    \item [\ding{184}] \textbf{Parameter Sanitization (SP)} validates the input parameters to ensure they conform to the system call’s expectations. It avoids unexpected execution errors caused by unreasonable input parameters. More importantly, parameter sanitization can sanitize inputs provided by untrusted threads to prevent vulnerabilities from being triggered by malicious inputs (e.g., confused deputy attack~\cite{hardy1988confused,DBLP:conf/ndss/MachiryGSSSWBCK17,DBLP:conf/uss/SuciuMSS20}). This typically involves verifying that pointer parameters reference legal memory regions and imposing proper range constraints on the values of sensitive non-pointer parameters.\looseness=-1
\end{itemize}

\subsection{Comparison with General-purpose OSs}
\label{OS:compariosn}
Note that these security protections are widely adopted in general-purpose operating systems, albeit with different implementations. For RTOSs, these differences stem from limited hardware resources and distinct software architectures. For instance, general-purpose operating systems rely on MMUs for virtual memory management, whereas RTOSs typically use MPUs to enforce access control within a single address space.
Additionally, the differences in software architectures lead to variations in security implementations. The existing architectures of RTOSs often limit the direct application of security protections used in general-purpose OSs. For example, Unix-like systems abstract kernel objects into file descriptors, enhancing security by preventing users from accessing kernel object addresses directly. In contrast, modern RTOSs often expose kernel object pointers to users. Applying file descriptor-like mechanisms to RTOSs would likely introduce memory overhead, degrade performance, and create interface incompatibilities. These limitations highlight the need for tailored security protections in RTOSs. However, their practical security effectiveness remains largely unexamined, warranting a comprehensive assessment of their security guarantees.

\section{Motivation and Key Insights}
\label{sec:understanding-security-protections}






\subsection{Motivation}

\paragraph{Possible Attacks against RTOSs}
\label{lab:attack-surface-analysis}
We conduct an in-depth analysis of the attack surface associated with the three security protections. We discuss possible attacks that could occur in the absence of each protection mechanism, as follows:


\begin{itemize}
    \item \textbf{Lack of Privilege Separation.} If privilege separation is disabled by the RTOS developer (or equivalently voided by attacks), both the kernel and the user threads operate at the privileged level. This presents a significant risk, as a compromised user thread can easily take control of the entire system through control flow attacks, such as Return-Oriented Programming (ROP)~\cite{DBLP:conf/uss/CarliniW14, DBLP:conf/trustcom/WeidlerBMAWCKWW17}. Moreover, with privileged execution, a compromised user thread can disable the MPU, effectively bypassing all memory access control policies.\looseness=-1
    
    \item \textbf{Lack of Memory Access Control.} Without memory access control, any thread can freely access system memory. Even if privilege separation is enforced, a compromised user thread can still read or manipulate kernel data without restriction. Additionally, in such a scenario, both code and data regions remain writable and executable, allowing attackers to inject and execute arbitrary code. \looseness=-1
    \item \textbf{Lack of Parameter Sanitization.} Without parameter sanitization, a compromised user thread can manipulate system call parameters to escalate privileges or gain arbitrary memory access~\cite{DBLP:conf/apsys/ChenMWZZK11,optee-advisories,DBLP:journals/csur/PintoS19,DBLP:conf/ccs/BulckOMAGP19,DBLP:conf/ccs/WangZHZGBLGC23, DBLP:conf/uss/LeeJJKCCKPK17}. For instance, since RTOSs often expose kernel object pointers as system call parameters, an attacker can pass a malicious pointer referencing sensitive data. If the system call lacks proper validation, the kernel may dereference this pointer, granting unauthorized privileged access to critical data. 
\end{itemize}



\begin{table}
    \centering
    \scriptsize
    \aboverulesep=2pt
    \belowrulesep=2pt
    \setlength{\tabcolsep}{2pt}
    \setlength{\aboverulesep}{2pt}
\caption{Protection Implemented in Various RTOSs}

\label{tab:protections-rtos}
\begin{tabular}{@{}cccc@{}}
\toprule
  & \multicolumn{2}{c}{\textbf{Hardware}}             & \textbf{Software} \\ \cmidrule(lr){2-3} \cmidrule(lr){4-4}
\multicolumn{1}{l}{} & \thead{Privilege \\ Separation (\ding{182})} & \thead{Memory Access \\ Control (\ding{183})} & \thead{Parameter Sanitization (\ding{184})}
\\ \hline

FreeRTOS & \tickYes                   & \tickYes                   & \tickNo \\ 
LiteOS-M      & \tickNo & \tickYes & \tickNo                           \\
Mbed OS      & \tickNo & \tickYes & \tickNo                           \\ 
\rowcolor{blue!20}
ThreadX      & \tickYes                   & \tickYes                   & \tickYes                   \\ \rowcolor{blue!20}
Zephyr OS    & \tickYes                   & \tickYes                   & \tickYes                   \\

\bottomrule
\end{tabular}
\vspace{-5mm}
\end{table}

\paragraph{Security Analysis of State-of-the-art RTOSs}
\label{para:vulnerabilities-in-mainstream-RTOSs}
As discussed above, each security protection is essential for safeguarding RTOSs against potential attacks. To evaluate their implementation in state-of-the-art RTOSs, we conducted a thorough analysis of both the source code and technical documentation. To be more specific, we reviewed their documentation, inspected the implementations, and designed PoC attacks to validate the findings. The results are presented in \autoref{tab:protections-rtos}. 
It can be observed that many RTOSs have missed at least one security protection, leaving them vulnerable to various attacks as discussed.

For example, FreeRTOS utilizes system calls with insufficient pointer parameter sanitization, which introduces vulnerabilities exploitable for privilege escalation and arbitrary memory access. We find all systems supporting FreeRTOS suffer from our identified vulnerabilities and CVE-2024-28xxx (anonymized) has been generated for the issue (We provide detailed examples for readers who are interested in Appendix~\ref{appendix-a:weakpointercheck}).  

Among the RTOSs we investigated, Zephyr OS and ThreadX stand out for their complete security protections including privilege separation, memory access control, and parameter sanitization. However, while their implemented privilege separation and memory access control are similar, significant differences exist in their parameter sanitization of kernel object pointer parameters.
Zephyr OS employs a cautious strategy by storing the kernel object pointer addresses and semantics, such as types and states, of all created kernel objects within kernel memory. This practice helps sanitize kernel object pointer parameters by verifying that the pointer corresponds to a valid address and matches the expected semantics. Similarly, general-purpose operating systems like Linux manage a mapping of kernel object pointers to valid index variables to ensure secure access.

However, ThreadX adopts a more efficient strategy. 
This is evidenced by our experimental analysis, which shows that the parameter sanitization in ThreadX incurs an overhead of approximately 500 CPU cycles, compared to around 1000 CPU cycles in Zephyr OS, with both measurements conducted on the same board. We then ask a few questions: \textit{How does ThreadX perform parameter sanitization? Are those parameter sanitization mechanisms secure or not? If not, how could we explore them?} As such, this work seeks to address these questions by evaluating the implementations and devising proof-of-concept attacks to raise awareness.

\subsection{Key Insights} 
\label{subsec:insight}

 After a deeper investigation, this efficiency is mainly due to ThreadX's employment of a more streamlined parameter sanitization approach. As illustrated in \autoref{fig:parameter-checking-flow}, the sanitization process comprises two essential components:

\begin{figure}[ht]
\centering
\includegraphics[width=\linewidth]{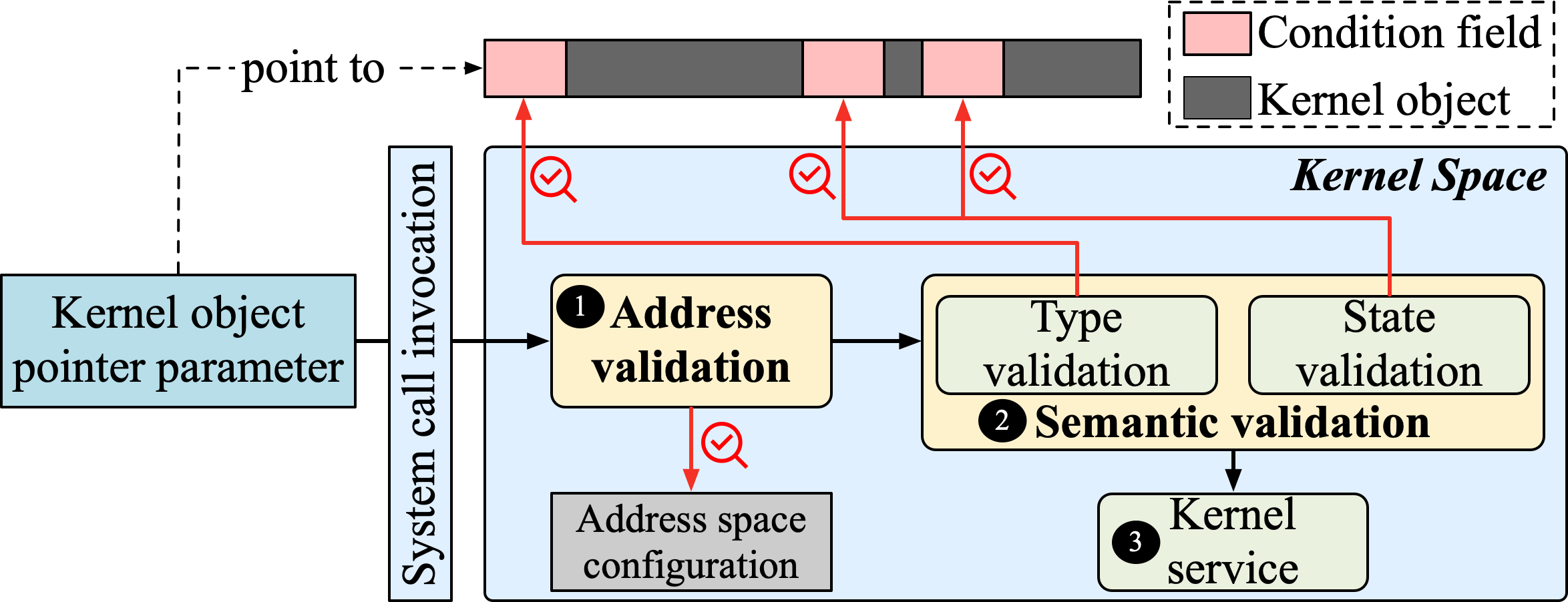}
\vspace{-3mm}
\caption{Overview of Parameter Sanitization in ThreadX}
\label{fig:parameter-checking-flow}
\vspace{-5mm}
\end{figure}

\begin{itemize}
    \item \textbf{Address Validation (\ding{182})} 
verifies whether the input pointer falls within an expected address range. For kernel object pointers, it must fall outside the isolated memory of the currently running thread (instead of maintaining and checking it against all valid kernel object pointers). As kernel memory lies beyond the scope of currently running thread memory, ThreadX will prevent the running threads from directly altering contents within kernel memory. This modification enhances performance by eliminating the exhaustive checking of the kernel object pointer addresses. \looseness=-1

\item \textbf{Semantic Validation (\ding{183})} examines certain fields (referred to as \textit{condition fields}) of the pointed kernel object to ensure they meet semantic requirements, instead of searching for maintained information to acquire semantics. 
The checks can be one of the following two types:   
{(i) The first field of a kernel object is always scrutinized as it serves as the type identifier of the kernel object. The validation succeeds if the first field aligns with a predefined type ID (i.e., Type Validation). (ii) Other specific fields that represent the states of the kernel object are checked based on the distinct requirements of each system call (State Validation). Note that each system call must check at least one condition field of the pointed kernel object. 
These validations ensure that threads cannot arbitrarily redirect the input kernel object pointers within the address range specified by address validation but are restricted to legitimate kernel objects only. This modification enhances performance by eliminating the exhaustive maintaining and checking of kernel object pointer semantics.}

\end{itemize}

\begin{figure}[ht]
\centering
\vspace{-2mm}
\includegraphics[width=\linewidth]{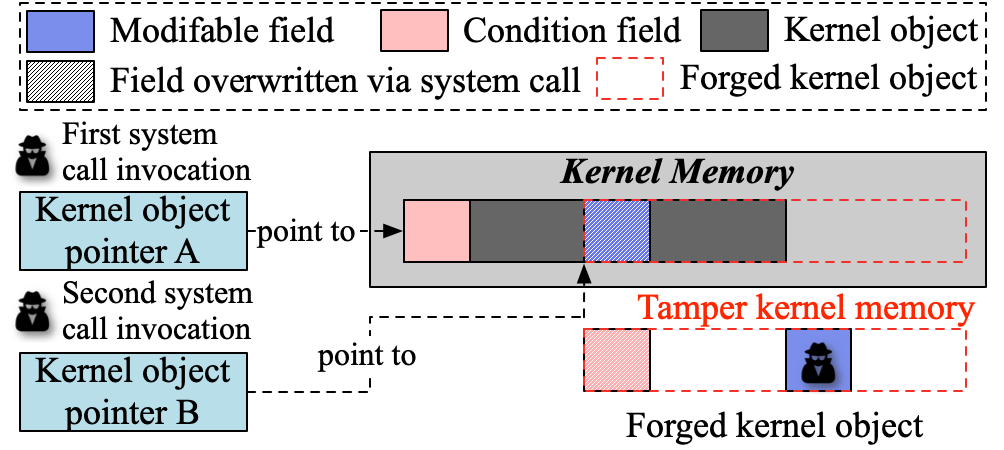}
\vspace{-3mm}
\caption{Bypass Parameter Sanitization of ThreadX}
\label{fig:bypass-parameter-sanitization-of-ThreadX}
\vspace{-5mm}
\end{figure}

While ThreadX improves its performance by avoiding exhaustive checking of kernel object pointer addresses and semantics, we find the parameter sanitization on kernel object pointers to be insufficient. Specifically, we have two observations (O): 
\begin{itemize}
    \item \textbf{(O-I):} The address validation requires that the input kernel object pointer falls outside the isolated memory of the currently running thread. In other words, any kernel object pointers (originally designed to access kernel memory, not the isolated memory of the running thread) meet this requirement.
    \item  \textbf{(O-II):}  We notice that some system calls allow threads to modify certain fields of a kernel object with other parameters (referred to as \textit{modifiable fields}). This means that a user thread can set the modifiable fields of a valid kernel object to certain values, making these fields masquerade as the condition fields required to satisfy the semantic validation of another system call. This is because when ThreadX performs the check, it needs to locate the condition fields through the kernel object pointer parameter, which can be manipulated by a user thread.
\end{itemize}
These two observations motivate us to devise a method to bypass the parameter sanitization process for kernel object pointer parameters by invoking a carefully selected system call twice (or two distinct system calls): As shown in \autoref{fig:bypass-parameter-sanitization-of-ThreadX}, if the modifiable fields of a kernel object (pointed to by kernel object pointer A) align with the condition field of another kernel object (pointed to by kernel object pointer B), a user thread can set the modifiable fields of the first kernel object to a specific value (by invoking the system call the first time) that satisfy the semantic validation when the system call is invoked a second time. Consequently, we construct a ``virtual'' kernel object (called \textit{forged kernel object}) that overlaps the modified kernel object in kernel memory. This forged kernel object can masquerade as a valid kernel object to bypass the parameter sanitization process. Firstly, the kernel object pointer, which references the forged kernel object, can pass the address validation based on observation O-I. Meanwhile, having the same condition fields as system call B, the forged kernel object can pass the semantic validation based on observation O-II.\looseness=-1

At this point, astute readers may realize that, beyond bypassing parameter sanitization, a malicious user thread can also tamper with kernel memory during the second invocation of the system call by modifying the modifiable fields of the forged kernel object. This capability motivates us to develop a more advanced attack to further explore the impact of this insecure implementation in ThreadX.\looseness=-1

\looseness=-1

\section{Kernel Object Masquerading Attack}

In this section, we explain how we leverage the observations discussed in \S\ref{subsec:insight} to conduct a powerful attack, which we have named the Kernel Object Masquerading (KOM) Attack. This attack involves continuously modifying specific fields within forged kernel objects to carefully selected values by invoking a sequence of system calls. By doing so, these forged objects can masquerade as valid kernel objects when passed to these system calls.
\looseness=-1

\subsection{Threat Model and Scope}
\label{sub:threat-model}

We define the threat model for exploiting the vulnerabilities in the parameter sanitization mechanism of ThreadX.
We make certain assumptions regarding both the victims and the attackers. For the victims, we assume that the target application operates with ThreadX and that both privilege separation and memory access control mechanisms are enforced.
Regarding the attackers, we assume that they can gain control of an unprivileged thread through a vulnerability in the target application, enabling them to invoke the system calls with arbitrary parameters. This assumption is reasonable and achievable, as numerous existing techniques, such as ROP, allow attackers to meet these conditions~\cite{DBLP:conf/ccs/CraneVSLLDSHSF15, DBLP:conf/asplos/JelesnianskiIJW23, exploitation-Returning-to-libc}. We adopt this threat model for the following stages of our attack. \looseness=-1

\subsection{Basic Idea}
\label{sub:basic-idea}

{In this attack, our goal is to access sensitive memory, typically with high privileges, using a carefully selected sequence of system calls targeting two initial kernel objects in memory: a malicious kernel object and an accomplice kernel object. These system calls create a set of forged kernel objects based on the malicious kernel object, allowing us to overwrite a pointer field of the accomplice kernel object and then leverage this pointer to achieve arbitrary memory access. We provide the detailed workflow of the KOM attack below. For clarity in our exposition, we simplify the semantic validation to focus solely on type validation (i.e., validating only the first field of the kernel object).
}

\begin{figure}[ht]
\vspace{-3mm}
\centering
\includegraphics[width=\linewidth]{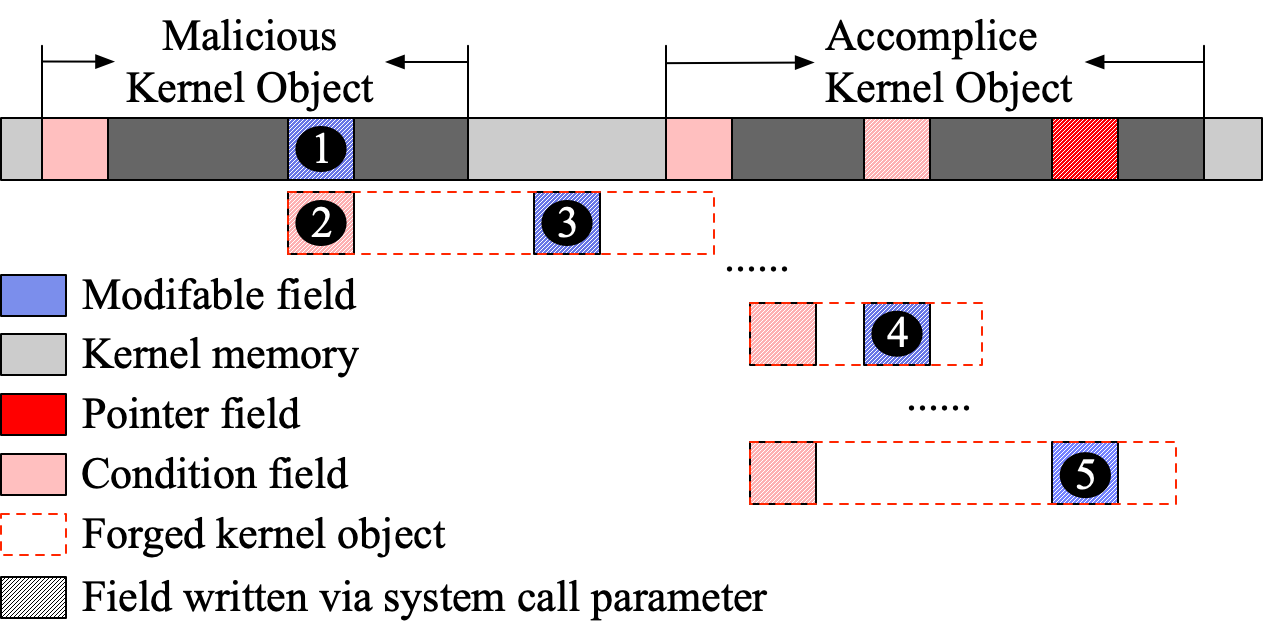}
\vspace{-5mm}
\caption{Workflow of KOM Attack}
\label{fig:ptr-attack}
\vspace{-5mm}
\end{figure}


\begin{enumerate}
\item {\em Creating/identifying \textit{malicious kernel object} and \textit{accomplice kernel object}}. These two kernel objects can be existing kernel objects or can be created through appropriate system calls. 
The malicious kernel object shall have at least one modifiable field. The accomplice kernel object shall contain a pointer that can be dereferenced.
    
\item {\em Creating initial forged kernel object}. We then invoke a system call to change a modifiable field of the malicious kernel object to a chosen kernel object type ID (\ding{182}). We fabricate a pointer so that it points to the selected kernel object type ID, and a forged kernel object referred to by the pointer is created in this way. This forged kernel object, which has a valid type ID (\ding{183}), shall have at least one modifiable field (\ding{184}).
    
\item {\em Creating sequences of forged kernel objects to overwrite fields of the accomplice kernel object}. We now invoke appropriate system calls with a pointer to the initial forged kernel object to modify its modifiable fields. It is important to note that different system calls may be used, as they can modify different fields of the kernel object.
Multiple forged kernel objects of different types may be created out of the initial forged kernel object. The new forged kernel objects can then be used to create other forged kernel objects and sequences of forged kernel objects can be created recursively in this way. 
We create multiple sequences of forged kernel objects so as to change the target pointer (\ding{186}) and other condition fields (\ding{185}) in the accomplice kernel object.

    
    
    
    
\item {\em Dereferencing pointer of accomplice kernel object}. After the target pointer is overwritten to point to sensitive data of interest and condition fields of the accomplice kernel object are changed accordingly, the pointer can then be dereferenced through a system call. For example, if the pointer points to the MPU control register~\cite{armv8-m-mpu,threadx-mpu-protection}, the MPU can then be disabled and an unprivileged thread can access the kernel memory now.

\end{enumerate}

Although the basic idea is straightforward, conducting KOM attacks is not trivial, as several scenarios must be considered during the attack process:

\begin{itemize}
\item ThreadX has a defined set of system calls, each capable of executing various memory operations on different types of kernel objects. How can we identify the vulnerable system calls and the specific fields (i.e., modifiable fields and condition fields) that can be exploited to construct forged kernel objects?
\item The malicious and accomplice kernel objects can be of any type. Moreover, the accomplice kernel object is located at a specific distance from the malicious kernel object in memory. How can we select appropriate malicious and accomplice kernel objects, considering their types and memory locations?
\item System calls that modify specific fields and dereference pointers may also validate one or more additional fields (referred to as condition fields). To successfully dereference the pointer, both the condition fields and the pointer must be modified accordingly. How can we design a sequence of system calls to construct suitable forged kernel objects, starting from the malicious kernel object, to overwrite the target memory locations (i.e., the pointer and condition fields in the accomplice kernel object)?
\end{itemize}

While manually addressing these challenges is feasible, it is labor-intensive and prone to errors. Therefore, we propose an automated method to efficiently tackle these issues.

\section{Automated Method for Mounting the KOM Attack}

In this section, we present an automated solution for mounting the KOM attack, including three phases as follows:
\begin{itemize}

    \item [(\textbf{P1})] \textbf{Identifying Vulnerable System Calls (\S\ref{subsec:syscall-identification}):} We introduce a method based on dynamic symbolic execution to filter out the vulnerable system calls. By symbolically executing each system call, we can determine whether it is vulnerable through analyzing its memory modification capability of kernel objects as well as the corresponding path constraints, corresponding to the modifiable fields and condition fields, respectively. 
    \item [(\textbf{P2})] \textbf{Targeting Initial Kernel Objects (\S\ref{subsec:attack-preparation}):} Based on the definition of malicious kernel and accomplice kernel objects, we analyze all types of kernel objects used in the system calls so that we can determine appropriate types of these kernel objects and their memory locations.
    \item [(\textbf{P3})] \textbf{Generating System Calls Sequences (\S\ref{subsec:syscall-selection}):} We design a depth-first search (DFS) based algorithm to find an optimal sequence of system calls to modify the desired fields of the accomplice kernel object. A generated sequence of system calls can leverage their diverse memory modification capabilities to sequentially construct forged kernel objects at various memory locations, and ultimately modify a target memory location with the last created forged kernel object.
    
\end{itemize}

\begin{figure}[ht]
\centering
\vspace{-3mm}
\includegraphics[width=\linewidth]{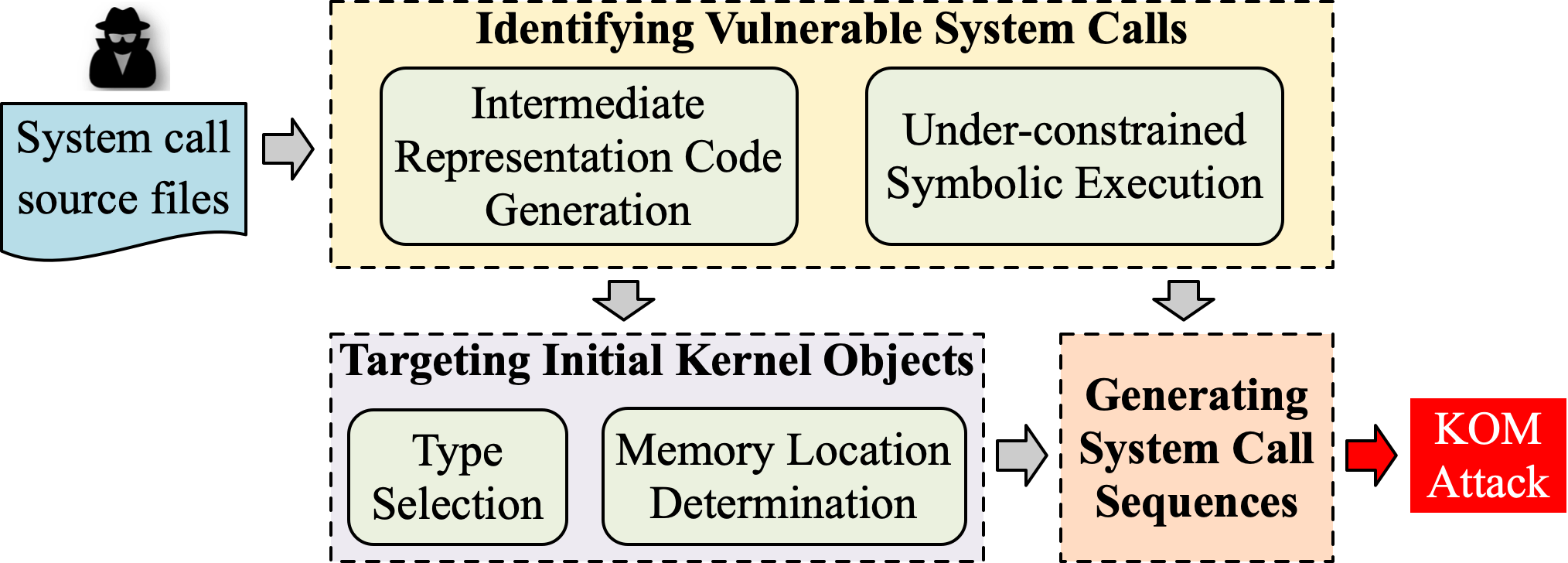}
\caption{Automated Method for Mounting the KOM Attack}
\label{fig:kernel-object-pointer-description}
\vspace{-5mm}
\end{figure}

\subsection{Identifying Vulnerable System Calls}
\label{subsec:syscall-identification}

We leverage dynamic symbolic execution to identify the ones that can be used to create forged kernel objects among all system calls of ThreadX. 
To be more specific, (i) We generate the Intermediate Representation (IR) code for each system call from the source code as input for the symbolic execution. (ii) We customize an under-constrained symbolic execution engine to simulate the execution of each individual system call based on its IR code. \looseness=-1

\begin{figure}[ht]
\centering
\includegraphics[width=\linewidth]{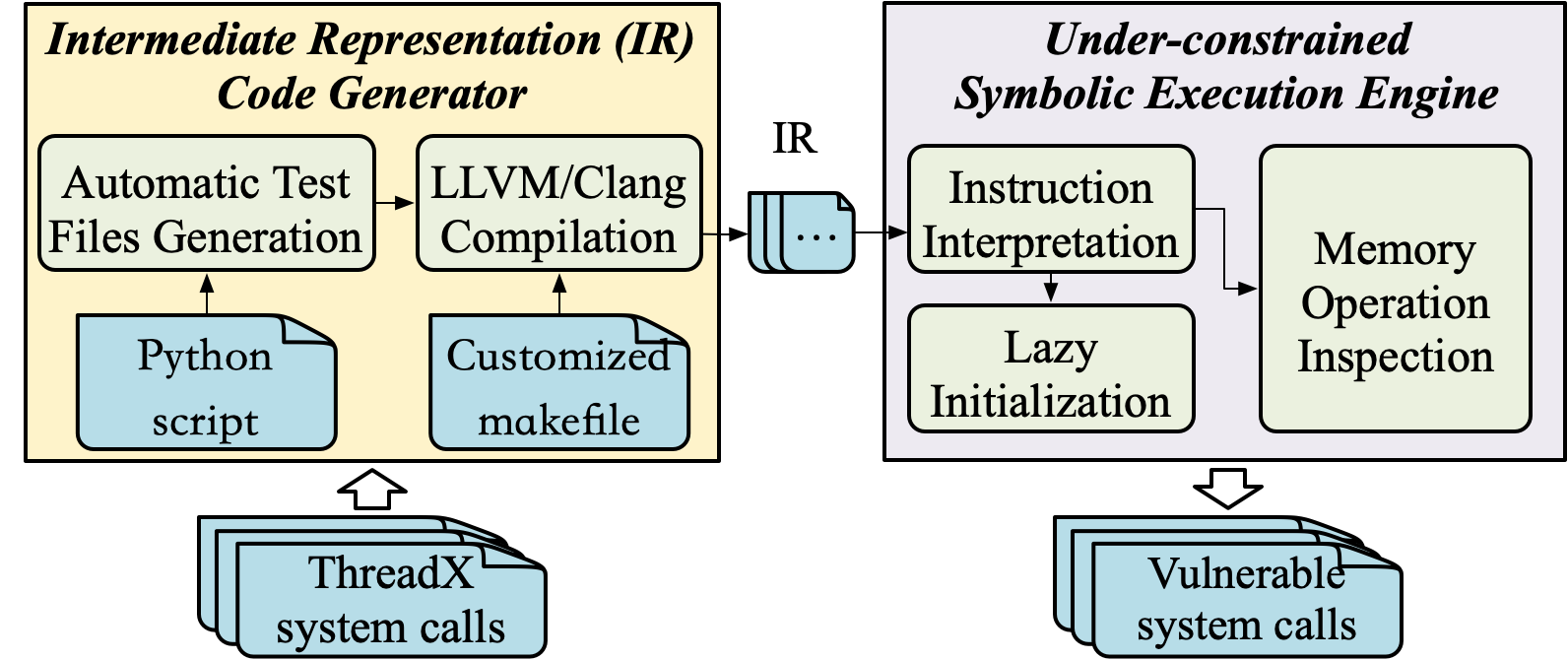}
\caption{Vulnerable System Calls Identification Flow}

\label{fig:feature-extraction}
\vspace{-5mm}
\end{figure}

\paragraph{Step I - Intermediate Representation Code Generation} 
As shown in \autoref{fig:feature-extraction}, we write a Python script to parse all declarations of ThreadX's system calls and generate test files for each system call. Each test file contains an entry point that would symbolize all parameters of the system call and invoke it subsequently. To compile each system call to IR individually, we customize a makefile to handle the compile-time dependencies.

\paragraph{Step II - Under-constrained Symbolic Execution}
To analyze the system calls, we customize an under-constrained symbolic execution engine~\cite{DBLP:conf/usenix/RamosE16} to symbolically execute each individual system call. Unlike standard symbolic execution, which requires a complete program and fully specified inputs, under-constrained symbolic execution allows analysis to proceed even when some parts of the program or its inputs are missing. This technique is useful for analyzing individual functions like system calls without needing the entire program context. Moreover, it reduces the execution of code unrelated to these calls, thereby alleviating the burden on symbolic execution and enabling us to focus on the memory operations on kernel objects for each system call.

However, under-constrained symbolic execution introduces challenges such as context loss due to under-constrained variables (e.g., pointer parameters and uninitialized global variables). To mitigate this issue, we first manually initialize all uninitialized global variables in the test files by reviewing the source code. Then, we leverage lazy initialization~\cite{DBLP:conf/tacas/KhurshidPV03, DBLP:journals/ase/DengLR12} to resolve the pointer parameters, which can be identified during the instruction interpretation. 



After that, we inspect the memory operations of each system call during symbolic execution and filter out vulnerable system calls. To create forged kernel objects, we attempt to find system calls that can modify specific fields (i.e., modifiable fields) to desired values (e.g., type IDs) within the kernel objects. The ability to modify these fields allows us not only to forge type IDs but also to overwrite other fields (e.g., the pointer or condition fields in the accomplice kernel object). Therefore, we concentrate on all memory write operations of each system call to identify those that can be influenced by parameters so that an attacker could exploit them.

To be more specific, when a memory write operation occurs, we extract the symbolic expressions related to the memory address and the data to be written. We then examine the relevant memory write operations based on the following criteria: (i) The memory address should point to a field inside the kernel object. (ii) The data can be written to arbitrary values through the parameters.

To evaluate these criteria, we leverage taint-style analysis and a constraint solver to determine if the memory write operation can simultaneously satisfy both criteria. For criterion (i), we first obtain the pointed memory object by resolving the memory address expression in our symbolic execution engine. We then verify if the pointed memory object is indeed the kernel object operated on by the system call. To identify the kernel object among all memory objects, we taint the kernel object during the lazy initialization for the kernel object pointer parameter.

For criterion (ii), we taint all parameters before invoking the system call in the test file. Thus, we can check if the data expression is tainted during a memory write operation. It is important to note that even if the data is tainted by the parameters, there is no guarantee that this memory write operation can write arbitrary values to specific fields within the kernel object. For example, the parameters may be constrained by the path constraints, preventing the corresponding data from being modified to arbitrary values. Therefore, we utilize the integrated solver in our symbolic execution engine to test if the data expression can equal to desired values for condition fields (e.g., type IDs of each kernel object type). 

Next, we identify all modifiable fields and condition fields within the kernel object based on these memory write operations. These modifiable fields and condition fields play a crucial role in the system call sequence generation for our attack \S\ref{subsec:syscall-selection}.
For modifiable fields, we obtain their relative locations in the kernel object by resolving the memory address expressions. However, automatically identifying the condition fields is challenging from the path constraints. Since each path constraint may involve complex expressions with arithmetic or logic operations, analyzing all path constraints requires a parser capable of parsing diverse expressions and resolving pointer expressions if needed. Consequently, significant engineering effort is required to determine the condition fields from path constraints. Fortunately, our evaluation (as described in \S\ref{subsec:experiment-results}) reveals that the number of path constraints is limited (not exceeding six). This allows us to record all path constraints and manually deduce the condition fields for each system call.\looseness=-1




\subsection{Targeting the Initial Kernel Objects.}
\label{subsec:attack-preparation}
In this phase, we target the initial kernel objects (i.e., malicious kernel object and accomplice kernel object) in memory on the victim system. To be more specific: (i) We have to select exploitable kernel object types according to the vulnerable system calls since each type of kernel object is only applicable to a subset of all system calls.
(ii) We need to determine the memory locations of these two kernel objects to measure the distance between them, guiding our creation of forged kernel objects.

\paragraph{Type Selection} 
Both malicious and accomplice kernel objects shall meet necessary requirements for conducting KOM attack as illustrated in \S\ref{sub:basic-idea}. Based on the modifiable fields of the vulnerable system calls we identified earlier, we can determine the qualified types for malicious kernel objects. For accomplice kernel objects, careful consideration is required because the exploitability of the pointers varies depending on how they are dereferenced.

We prioritize pointers with “fully controllable pointer dereferencing,” which allows for arbitrary memory write or read operations. When dereferencing these pointers for a memory write operation, the data to be written should either be assigned through a parameter of the system call or a member variable within the kernel object, both of which can be controlled by an attacker. The pointers used in memory read operations are similar. Nevertheless, it should be acknowledged that other pointers (e.g., those that are dereferenced to write a fixed value), can still be carefully exploited to cause significant impact. For example, a selected fixed value can be used to overwrite the MPU control register, disabling the ENABLE bit and turning off memory access control protection. To identify suitable types for accomplice kernel objects, we use our dynamic symbolic execution engine in a similar manner by analyzing the pointer dereferencing operations of each system call.\looseness=-1

\paragraph{Memory Location Determination} 
Generally, we can determine the memory locations of malicious and accomplice kernel objects in two ways. The first method is to obtain the addresses of the kernel objects in the kernel object creation. In ThreadX, we can invoke specific system calls to create a desired type of kernel object at runtime. For example, a timer kernel object can be created by sequentially calling \texttt{module\_object\_allocate} and \texttt{timer\_create}, which allocate a memory chunk for the kernel object and initialize it, respectively. The \texttt{module\_object\_allocate} function returns the address of the created kernel object.


Alternatively, we can select the desired types of malicious and accomplice kernel objects from the pool of existing kernel objects. This method is applicable when we are restricted from invoking system calls for kernel object creation (e.g., the developer removes these system calls). In this scenario, we can determine the memory locations of these kernel objects by accessing the kernel object pointers stored in memory (e.g., on the stack), which can be accessed by a user thread. In this way, reverse engineering is often required to aid the process of identifying these pointers and determining the types of the corresponding kernel objects.

\label{subsub:system-call-chaining}
\subsection{Generating System Calls Sequences}
\label{subsec:syscall-selection}
In theory, selecting an optimal sequence of system calls for the KOM attack is an NP-hard problem. Fortunately, the number of vulnerable system calls is limited, which drives us to design an algorithm that uses depth-first search (DFS) based on the identified  vulnerable system calls. 

In this algorithm, we assume that the malicious kernel object and the accomplice kernel object have been identified as outlined in \S\ref{subsec:syscall-identification}. For each vulnerable system call, we maintain a record that details the relative positions of their modifiable fields and condition fields within the kernel object. Let $S$ denote the start address of the malicious kernel object, while $D$ represents the target memory location (which could be a pointer field or a condition field within the accomplice kernel object).
The algorithm proceeds as follows:
\begin{enumerate}[1.]
    \item \textbf{Variable Initialization.} Initialize the \textit{current kernel object pointer} to $S$. Set the \textit{current sequence} of system calls as an empty set. All sequences that provide feasible solutions will be stored in a set called \textit{solutions}.
    \item \textbf{System Call Enumeration.} Enumerate a system call from the set of all vulnerable system calls. Our enumeration iterates over each system call to explore all possible sequence combinations: (i) If it is the first system call, select one that operates on the specific type of the malicious kernel object, then proceed to Step 3. (ii) For subsequent system calls, select any system call that meets the condition: All condition fields of this system call, based on the \textit{current kernel object pointer}, must align with the modifiable fields of the previous system call, based on the previous kernel object pointer.
    
    \item \textbf{Modifiable Field Enumeration.} Enumerate each modifiable field of the current system call, updating the current kernel object pointer. Let $M_i$ denote the offset of the modifiable field. Then, evaluate the following conditions:
    \begin{enumerate}[i.]
        \item If $S + M_i == D$, the current sequence represents a feasible solution. Add this solution to \textit{solutions} and proceed to enumerate the next modifiable field.
        \item If $S + M_i > D$, the current sequence is not a feasible solution. Continue enumerating the next modifiable field. If all fields have been enumerated, return to Step 2.
        \item If $S + M_i < D$, update the current memory location to $S = S + M_i$. Set $S$ as the new kernel object pointer (i.e., the start address of the new forged kernel object). Add this system call to the \textit{current sequence} and return to Step 2 to enumerate the next system call.
    \end{enumerate}
    \item \textbf{Optimal Solution Selection.} From all feasible solutions in \textit{solutions}, select the one with the fewest number of system calls as the optimal solution.
\end{enumerate}

By employing this algorithm, we can generate appropriate sequences of system calls to execute the KOM attack, allowing us to overwrite the targeted fields of the accomplice kernel object.

\section{Evaluation}
\label{sec:Evaluation}
In this section, we address the following research questions (RQs) to comprehensively evaluate the KOM attacks.

\begin{itemize}
    \item [RQ1] {\textbf{What system calls can be used to perform KOM attacks? }}
    Out of all system calls, we aim to identify those (i.e., the vulnerable system calls) that an attacker can leverage to mount KOM attacks.
    


    \item [RQ2] \textbf{Can our attack be effective in various attack environments?}
In practice, the malicious kernel object and the accomplice kernel object may be separated by an uncertain distance (e.g., due to the unpredictability of dynamic memory allocation) in various attack environments. This uncertainty in distance may result in the absence of a solution to the aforementioned problem. Therefore, we further analyze the existence of solutions under different distances when conducting our attack.
    \item [RQ3] \textbf{How is the exploitability of each type of kernel object in KOM attack?} 
Given that selecting appropriate types for the malicious and accomplice kernel objects is critical for executing KOM attacks, we perform a comprehensive analysis of all kernel object types involved in the explored system calls.
    \item [RQ4]\textbf{How efficient is our symbolic execution?}
Since it is an exhausting process to identify system calls, we aim to evaluate the performance of our automated system call identification in terms of analysis speed and code coverage.\looseness=-1

    \item [RQ5] \textbf{What are the implications of our KOM attacks?}
    Considering that ThreadX has been pervasively deployed among various embedded platforms, we aim to understand whether real-world embedded platforms are vulnerable to KOM attacks. \looseness=-1

\end{itemize}

\subsection{Experiment Setup}

\paragraph{Target system calls}
Among all system calls of ThreadX, we selected 60 of them for our evaluation, filtering out those system calls that do not accept a kernel object pointer parameter (e.g., module\_allocate) as well as those tracing the system events used for debugging purposes. 

\paragraph{Symbolic Execution Engine}
We implemented our symbolic execution using an Ubuntu 20.04 server, equipped with an Intel(R) Xeon(R) E5-2620 v2 CPU and 64G RAM. 
To identify the system call pairs, we developed an under-constrained symbolic execution engine based on KLEE \cite{klee}. 
Specific optimizations are made for the RTOS environment. For instance, we used symbols to simulate operations related to hardware, such as defaulting to a symbolic value for all operations accessing peripherals. For loops, we set a loop threshold, and any loop exceeding this threshold would exit. 

\paragraph{Target Platforms}
To evaluate the effectiveness of the KOM attacks across different platforms, we selected several development environments, including actual development versions and QEMU simulations. These platforms encompass popular Cortex-M processors, including Cortex-M33 and Cortex-M4.



\begin{table}[!t]
    \centering
    \scriptsize
    \aboverulesep=0pt
    \belowrulesep=0pt
    \setlength{\tabcolsep}{1pt}
    \setlength{\aboverulesep}{0pt}
\setlength{\belowrulesep}{0pt}
\caption{\small Results of Vulnerable System Calls Identification. $M_{\text{1}}$ indicates the number of fields that can be modified. $M_{\text{2}}$ indicates the number of fields that can be affected by the parameters. $M_{\text{3}}$ indicates the number of fields that can be written to forged IDs. \textit{$C_{\text{max}}$/$C_{\text{min}}$} indicates the maximum/minimum number of condition fields. \textit{Status} indicates the system call exits normally (\textit{N}) or abnormally (\textit{A}) during the symbolic execution. \textit{\#Path} represents the number of paths explored by symbolic execution. \textit{\#Ins} indicates the number of instructions that the symbolic execution engine executes.}
\label{tab:system-call-pairs-for-constructing-forged-kernel-objects}
\begin{tabular}{cccccclll}
\hline
\multicolumn{5}{c}{\textbf{Identification Result}} &
  \multicolumn{4}{c}{\textbf{Performance Metrics}} \\ \hline
\textbf{System Call} &
  \textbf{\#$M_1$} &
  \textbf{\#$M_2$} &
  \textbf{\#$M_3$} &
  \textbf{\#$C_{\text{max}}$/$C_{\text{min}}$} &
  \textbf{Status} &
  \multicolumn{1}{c}{\textbf{Time}} &
  \multicolumn{1}{c}{\textbf{\#Path}} &
  \multicolumn{1}{c}{\textbf{\#Ins}} \\
block\_allocate             & 3  & 0  & 0 & 2/2  & N  & 23     & 236    & 14882    \\
block\_pool\_create         & 4  & 4  & 2 & 1/1  & N  & 9      & 88     & 5224     \\
block\_pool\_delete         & 0  & 0  & 0 & N/A  & N  & 47     & 567    & 26626    \\
block\_pool\_info\_get      & 0  & 0  & 0 & N/A  & N  & 12     & 236    & 8384     \\
block\_pool\_prioritize     & 0  & 0  & 0 & N/A  & N  & 6      & 57     & 3145     \\
block\_release              & 0  & 0  & 0 & N/A  & N  & 1      & 8      & 296      \\
byte\_allocate              & 0  & 0  & 0 & N/A  & N  & 9      & 60     & 4532     \\
byte\_pool\_create          & 6  & 6  & 2 & 1/1  & N  & 5      & 52     & 3215     \\
byte\_pool\_delete          & 0  & 0  & 0 & N/A  & N  & 49     & 567    & 26626    \\
byte\_pool\_info\_get       & 0  & 0  & 0 & N/A  & N  & 11     & 236    & 8384     \\
byte\_pool\_prioritize      & 0  & 0  & 0 & N/A  & N  & 6      & 57     & 3145     \\
byte\_release               & 0  & 0  & 0 & N/A  & N  & 34     & 209    & 11836    \\
event\_flags\_create        & 1  & 1  & 0 & 1/1  & N  & 5      & 34     & 2546     \\
event\_flags\_delete        & 0  & 0  & 0 & N/A  & N  & 49     & 567    & 26626    \\
event\_flags\_get           & 3  & 1  & 1 & 6/4  & N  & 124    & 957    & 68965    \\
event\_flags\_info\_get     & 0  & 0  & 0 & N/A  & N  & 7      & 134    & 4697     \\
event\_flags\_set           & 3  & 0  & 0 & 27/1 & N  & 403    & 6679   & 239501   \\
event\_flags\_set\_notify   & 2  & 0  & 0 & 1/1  & N  & 2      & 14     & 729      \\
mutex\_create               & 1  & 1  & 0 & 1/1  & N  & 8      & 58     & 4736     \\
mutex\_delete               & 3  & 0  & 0 & 4/3  & N  & 17,987 & 158975 & 9276099  \\
mutex\_get                  & 3  & 0  & 0 & 6/3  & N  & 540    & 5125   & 326769   \\
mutex\_info\_get            & 0  & 0  & 0 & N/A  & N  & 11     & 236    & 8384     \\
mutex\_prioritize           & 0  & 0  & 0 & N/A  & N  & 6      & 57     & 3145     \\
mutex\_put                  & 5  & 0  & 0 & 7/3  & N  & 33,648 & 289827 & 16714889 \\
queue\_create               & 8  & 8  & 2 & 1/1  & N  & 7      & 64     & 4426     \\
queue\_delete               & 0  & 0  & 0 & N/A  & N  & 48     & 567    & 26584    \\
queue\_flush                & 3  & 0  & 0 & 2/2  & N  & 45     & 570    & 23542    \\
queue\_front\_send          & 3  & 0  & 0 & 4/3  & N  & 473    & 11190  & 341484   \\
queue\_info\_get            & 0  & 0  & 0 & N/A  & N  & 13     & 236    & 8384     \\
queue\_prioritize           & 0  & 0  & 0 & N/A  & N  & 7      & 57     & 3145     \\
queue\_receive              & 5  & 0  & 0 & 7/3  & N  & 484    & 5379   & 294694   \\
queue\_send                 & 5  & 0  & 0 & 5/3  & N  & 500    & 11295  & 344838   \\
queue\_send\_notify         & 2  & 0  & 0 & 1/1  & N  & 2      & 14     & 729      \\
semaphore\_ceiling\_put     & 3  & 0  & 0 & 3/3  & N  & 62     & 1493   & 44416    \\
semaphore\_create           & 2  & 2  & 1 & 1/1  & N  & 4      & 34     & 2720     \\
semaphore\_delete           & 0  & 0  & 0 & N/A  & N  & 48     & 567    & 26584    \\
semaphore\_get              & 2  & 0  & 0 & 3/2  & N  & 21     & 214    & 13489    \\
semaphore\_info\_get        & 0  & 0  & 0 & N/A  & N  & 6      & 134    & 4667     \\
semaphore\_prioritize       & 0  & 0  & 0 & N/A  & N  & 6      & 57     & 3145     \\
semaphore\_put              & 2  & 0  & 0 & 3/3  & N  & 58     & 1487   & 42228    \\
semaphore\_put\_notify      & 2  & 0  & 0 & 1/1  & N  & 2      & 14     & 729      \\
thread\_create              & 15 & 12 & 7 & 1/1  & N  & 461    & 4018   & 279885   \\
thread\_delete              & 0  & 0  & 0 & N/A  & N  & 3      & 26     & 2439     \\
thread\_entry\_exit\_notify & 0  & 0  & 0 & N/A  & N  & 2      & 17     & 797      \\
thread\_info\_get           & 0  & 0  & 0 & N/A  & N  & 41     & 824    & 30026    \\
thread\_preemption\_change  & 2  & 2  & 2 & 5/3  & N  & 4      & 80     & 3273     \\
thread\_priority\_change    & 6  & 4  & 0 & 7/3  & A & N/A    & N/A    & N/A      \\
thread\_relinquish          & 0  & 0  & 0 & N/A  & N  & 1      & 10     & 386      \\
thread\_reset               & 1  & 0  & 0 & 2/2  & N  & 1      & 7      & 386      \\
thread\_resume              & 0  & 0  & 0 & N/A  & N  & 32     & 344    & 22128    \\
thread\_suspend             & 0  & 0  & 0 & N/A  & N  & 5      & 62     & 3597     \\
thread\_terminate           & 0  & 0  & 0 & N/A  & N  & 58     & 461    & 32214    \\
thread\_time\_slice\_change & 2  & 2  & 2 & 1/1  & N  & 1      & 23     & 840      \\
thread\_wait\_abort         & 0  & 0  & 0 & N/A  & N  & 46     & 512    & 29787    \\
timer\_activate             & 1  & 0  & 0 & 6/6  & N  & 2      & 28     & 1365     \\
timer\_change               & 2  & 2  & 2 & 2/2  & N  & 1      & 20     & 806      \\
timer\_create               & 7  & 5  & 3 & 2/1  & N  & 44     & 425    & 27154    \\
timer\_deactivate           & 1  & 0  & 0 & 6/4  & N  & 8      & 107    & 5103     \\
timer\_delete               & 0  & 0  & 0 & N/A  & N  & 5      & 26     & 2949     \\
timer\_info\_get            & 0  & 0  & 0 & N/A  & N  & 46     & 902    & 33275    \\ \hline
\end{tabular}
\label{tab:exploitable-system-calls-result}
\vspace{-3mm}
\end{table}

\subsection{Experiment Results}\label{subsec:experiment-results}



 


\paragraph{Vulnerable System Calls (RQ1)}\label{rq1}
As shown in \autoref{tab:exploitable-system-calls-result}, the results of vulnerable system calls identification reveal that 31 out of 60 system calls have the capability to modify fields within kernel objects, albeit with varying degrees of modification capabilities. Specifically, we identified 17 system calls that can only set fixed values to fields, while the remaining 13 system calls can alter fields based on parameters. Note that these system calls that set fields to fixed values can also be exploited to manipulate sensitive data through forged kernel objects, potentially leading to data corruption and kernel crashes.
 
For the 13 modifiable fields (denoted as $M_{\text{2}}$) that can be affected by the parameters, we identified that some of these fields can only be modified to restricted values rather than forged IDs or arbitrary ones. The results are shown in \autoref{tab:modifiable-fields-affected-by-parameters}. Among these system calls, 10 can modify their modifiable fields to forged IDs (denoted as $M_{\text{3}}$). In other words, these system calls can be used to create forged kernel objects in our attack chain. Moreover, we observed that only 6 system calls can write arbitrary values to their modifiable fields (denoted as $M_{\text{4}}$). They enable us to overwrite the specific fields (e.g., the pointer field in the accomplice kernel object) with any desired values. The results indicate that we can modify no more than three fields within a kernel object using a single system call. This suggests that ThreadX can enforce more than three condition fields within a kernel object to prevent attackers from constructing forged kernel objects.

Further analysis reveals that these restrictions stem from parameter sanitization or bitwise operations. Parameter sanitization imposes path constraint on the particular parameters. For instance, when creating a kernel object, a ``name'' field typically references a string in user memory via a pointer parameter. To prevent the pointer from referencing kernel memory, ThreadX enforces that this pointer parameter remains within the bounds of user memory through its system call wrapper (i.e., semantic validation). For bitwise operations, parameters are restricted by only allowing specific bits within a field to be modified rather than the entire field. Additionally, we found that a single parameter could simultaneously modify multiple fields within a kernel object, which we treat as identical due to their shared value. 

\looseness=-1

\autoref{tab:exploitable-system-calls-result} also shows the number of path constraints when modifying these modifiable fields. Note that we only focus on the path constraints associated with the kernel object so that we can locate the condition fields. Each path constraint is typically associated with a condition field. We found that each modifiable field must be accompanied by at least one condition field within the kernel object. This requirement exists because ThreadX must verify the type ID for system calls that operate on kernel objects, or check the size of the corresponding metadata for allocated memory in system calls that create kernel objects. Note that the more condition fields a system call has, the more challenging it is to construct a forged kernel object, as all requirements related to these condition fields must be met. As a result, the number of condition fields is an important reference when we construct forged kernel objects.\looseness=-1 

\begin{table}[!t]
    \centering
    \scriptsize
    \aboverulesep=0pt
    \belowrulesep=0pt
    \setlength{\tabcolsep}{2pt}
    \setlength{\aboverulesep}{0pt}
\setlength{\belowrulesep}{0pt}
\caption{Modifiable Fields Affected by Parameters. $M_{\text{4}}$ denotes the number of modifiable fields that can be written with arbitrary values. $R$ represents the number of modifiable fields with restricted values, and $I$ represents the number of modifiable fields with identical values.}
\begin{tabular}{ccccccc}
\hline
\textbf{Type} &
  \textbf{System Call} &
  \textbf{\#$M_{\text{2}}$} &
  \textbf{\#$M_{\text{3}}$} &
  \textbf{\#$M_{\text{4}}$} &
  \textbf{\#R} &
  \textbf{\#I} \\ \hline
BLOCK     & block\_pool\_create         & 4  & 2 & 0 & 3 & 0 \\
BYTE      & byte\_pool\_create          & 6  & 2 & 0 & 6 & 0 \\
EVENT     & event\_flags\_get           & 1  & 1 & 1 & 0 & 0 \\
EVENT     & event\_flags\_create        & 1  & 0 & 0 & 1 & 0 \\
MUTEX     & mutex\_create               & 1  & 0 & 0 & 1 & 0 \\
QUEUE     & queue\_create               & 8  & 2 & 0 & 8 & 0 \\
SEMAPHORE & semaphore\_create           & 2  & 1 & 1 & 1 & 0 \\
THREAD    & thread\_priority\_change    & 4  & 0 & 0 & 4 & 0 \\
THREAD    & thread\_preemption\_change  & 2  & 2 & 0 & 2 & 0 \\
THREAD    & thread\_time\_slice\_change & 2  & 2 & 1 & 0 & 1 \\
THREAD    & thread\_create              & 12 & 7 & 2 & 9 & 0 \\
TIMER     & timer\_create               & 5  & 3 & 3 & 2 & 0 \\
TIMER     & timer\_change               & 2  & 2 & 2 & 0 & 0 \\ \hline
\end{tabular}
\label{tab:modifiable-fields-affected-by-parameters}
\vspace{-3mm}
\end{table}


\paragraph{Attack Effectiveness among Various Attack Environments (RQ2)} \label{rq2}
We attempt to explore whether there is always a solution even when the target location is undetermined in memory. We observed an interesting property between the modifiable fields and the condition fields. That is, among all vulnerable system calls, several system calls have at least one modifiable field, which is with a higher memory address than all conditional fields within the kernel object. If we are able to modify all fields before such a modifiable field inside a forged kernel object, we can keep creating new forged kernel objects, allowing us to overwrite each field in memory, starting from the initial forged kernel object.\looseness=-1 

\begin{figure}[ht]
\centering
\includegraphics[width=\linewidth]{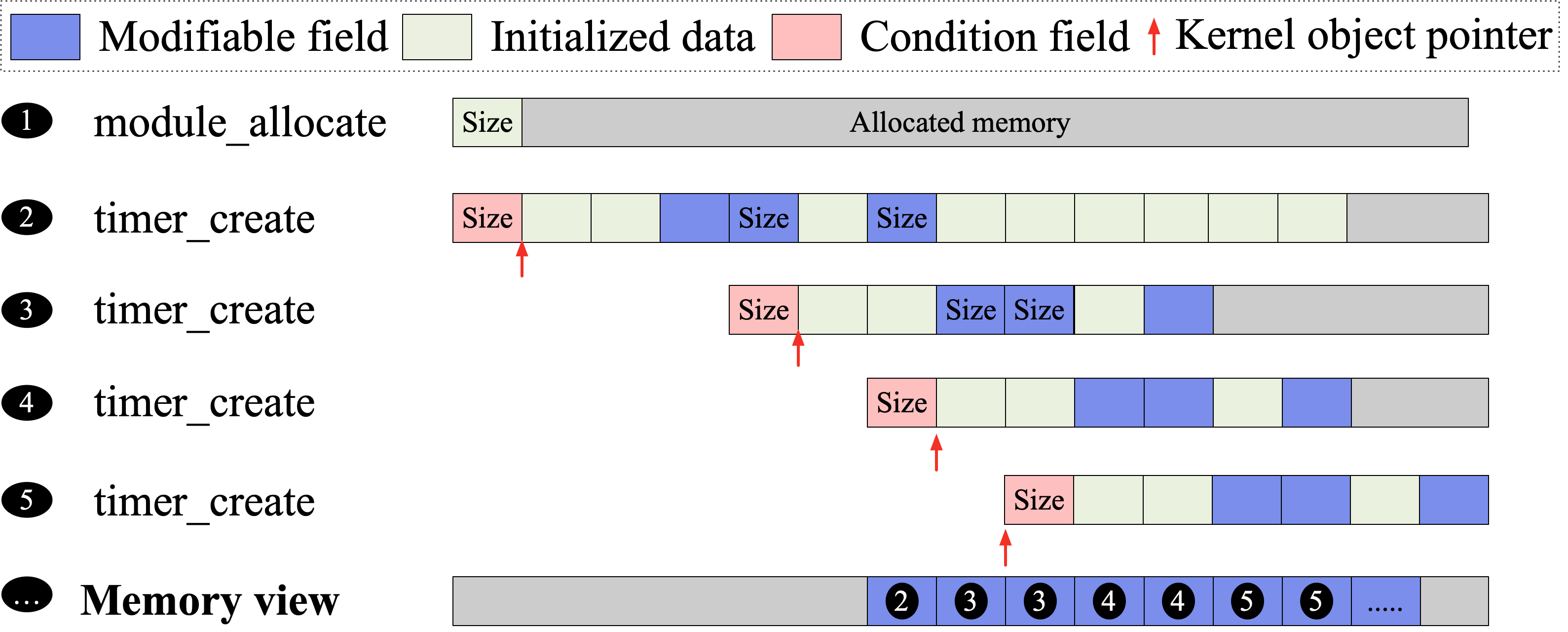}

\caption{Example of Successive Field Modification}
\label{fig:arbitrary-modification-within-memory-block}
\vspace{-3mm}
\end{figure}

For example, as shown in \autoref{tab:system-calls-for-successie-fields-modification}, we select two system calls that can be used to achieve this goal. As shown in \autoref{fig:arbitrary-modification-within-memory-block}, we create a malicious kernel object with the type of timer, then we can create a series of forged kernel objects separated by two fields and perform overwriting on every field in the memory. We first allocate a chunk of memory for a timer kernel object by invoking \texttt{module\_allocate} (\ding{182}). Then, we create a timer kernel object using \texttt{timer\_create} (\ding{183}). Simultaneously, we write the value of the size of a timer kernel object into two modifiable fields for creating two forged timer kernel objects. Afterward, we call \texttt{timer\_create} twice (\ding{184}, \ding{185}) in succession to modify the modifiable fields of these forged timer kernel objects. It can be observed that as long as we continue to assign a valid size when creating a forged timer kernel object, we can keep creating new forged timer kernel objects and write to higher memory addresses. The modifiable fields of these objects are continuously spread from lower to higher memory addresses, potentially covering more than seven consecutive fields in memory. However, in our experiments, we found that repeatedly calling \texttt{timer\_create} can cause noisy memory operations. This is because when the system creates the first timer, it initializes certain fields. If a subsequent call finds an existing timer kernel object, it may attempt to dereference associated fields, leading to a dereference error. Fortunately, we discovered that invoking \texttt{timer\_disable} can delete the existing timer object, causing the system to assume no timer kernel objects are present.

\begin{table}[!t]
    \centering
    \scriptsize
    \aboverulesep=2pt
    \belowrulesep=2pt
    \setlength{\tabcolsep}{15pt}
    \setlength{\aboverulesep}{2pt}
\caption{System Calls for Successive Fields Modification}
\label{tab:system-calls-for-successie-fields-modification}
\begin{tabular}{@{}cccc@{}}
\toprule
\textbf{System Call} & \textbf{\#Parameter} & \textbf{\#$M_{\text{4}}$} & \textbf{\#$C$} \\
module\_allocate          & 2                    & 0                          & 0                          \\
timer\_create             & 8                    & 3                          & 1                          \\ \bottomrule
\end{tabular}
\vspace{-3mm}
\end{table}

Based on this observation, we can conclude that as long as the accomplice kernel object is located at a higher address than the malicious kernel object, we can always have a solution that enables us to modify the necessary fields to achieve the intended overwrite.

\paragraph{Type for Malicious and Accomplice Kernel Objects (RQ3)}\label{rq3}
As shown in \autoref{tab:type-for-malicious-and-accomplice-kernel-object}, we list all suitable types for malicious and accomplice kernel objects along with the corresponding system calls according to our symbolic execution analysis result. Based on the findings in \autoref{tab:modifiable-fields-affected-by-parameters}, we can conclude that all types except \texttt{Mutex} are suitable for the malicious kernel object, since the corresponding system calls can modify fields within the kernel objects to desired values (i.e., forged type IDs). Moreover, the type of \texttt{Thread} should be given priority when selecting malicious kernel objects because it has more modifiable fields, allowing us to create more forged kernel objects simultaneously.

For accomplice kernel objects, we observed that all types are suitable for the accomplice kernel object, and all of them can achieve fully controllable pointer dereferencing. Further analysis of these system calls revealed that, among all cases of fully controllable pointer dereferencing, the data to be written is always assigned from a member variable within the kernel object rather than from the parameters of the system calls. The most significant instances of fully controllable pointer dereferencing occur during linked list operations. This is because ThreadX leverages several doubly linked lists to maintain kernel objects efficiently. These pointers (i.e., previous pointer and next pointer), which are inside the kernel objects as member variables, can be abused to achieve full pointer dereferencing during the list updating process.

\begin{figure}[ht]
\centering
\vspace{-3mm}
\includegraphics[width=\linewidth]{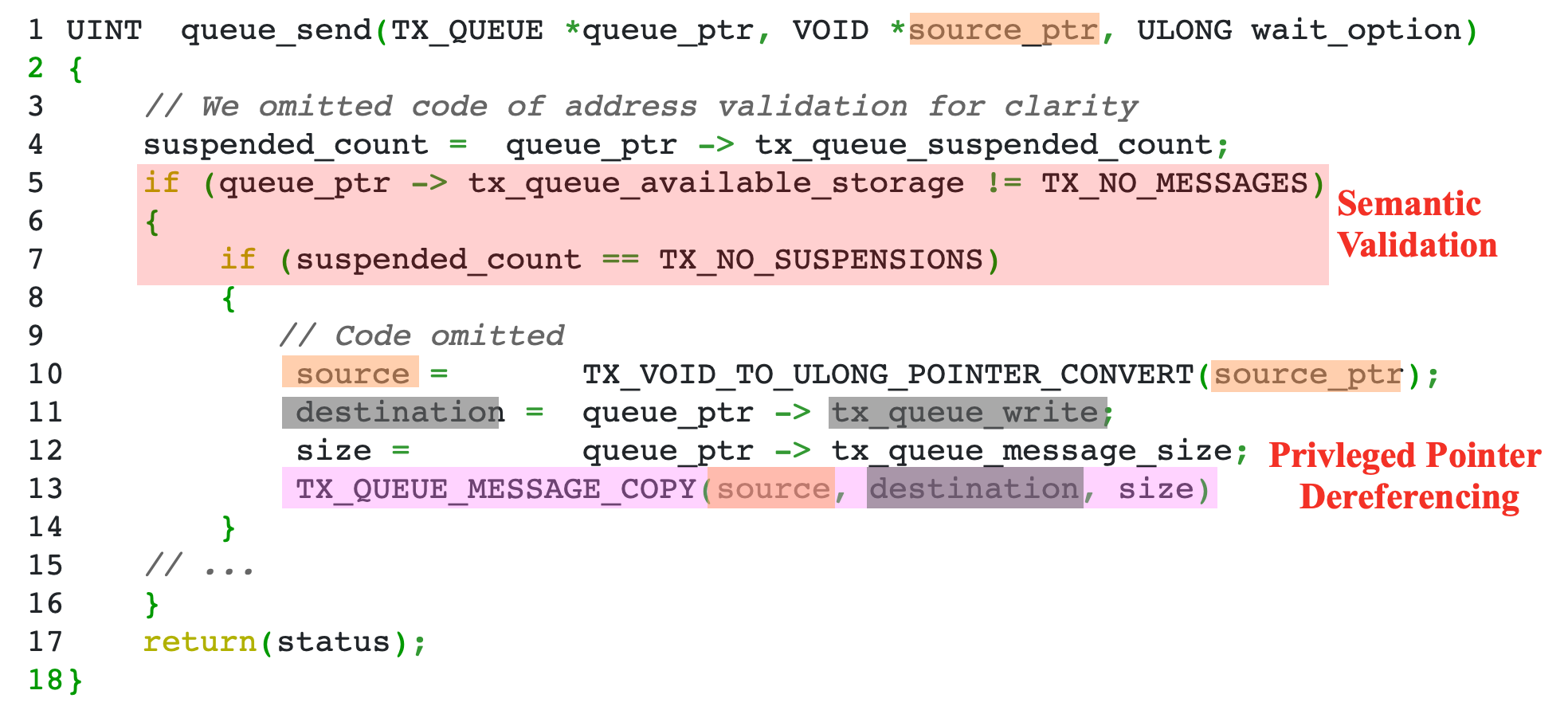}
\vspace{-4mm}
\caption{System Call for Fully Controllable Pointer Dereferencing}
\label{fig:code-queue-send}
\vspace{-2mm}
\end{figure}

Moreover, we found some system calls can also be exploited to overwrite an entire memory chunk rather than a specific data field, thereby increasing the security impact of KOM attacks.
For example, as shown in \autoref{fig:code-queue-send}, \texttt{queue\_send} can realize fully controllable pointer dereferencing on a queue kernel object (i.e., accomplice kernel object). Specifically, \texttt{queue\_send} call \texttt{TX\_QUEUE\_MESSAGE\_COPY} to copy data from the \texttt{source} to the \texttt{destination}, which is assigned by the parameter \texttt{source\_ptr} and the member variable \texttt{tx\_queue\_write} (i.e., privileged pointer). Besides, the \texttt{size} can also be controlled by an attacker as long as modifying the \texttt{tx\_queue\_message\_size} in the queue kernel object.

\begin{table}
    \centering
    \scriptsize
    \aboverulesep=0pt
    \belowrulesep=0pt
    \setlength{\tabcolsep}{2pt}
    \setlength{\aboverulesep}{0pt}
\setlength{\belowrulesep}{0pt}
\caption{Type for Malicious and Accomplice Kernel Object}
\centering
\begin{tabular}{@{}ccccc@{}}
\toprule
\textbf{Type} &
  \textbf{\begin{tabular}[c]{@{}c@{}}Malicious \\ Kernel Object\end{tabular}} &
  \textbf{\#$M_{\text{3}}$} &
  \textbf{\begin{tabular}[c]{@{}c@{}}Accomplice \\ Kernel Object\end{tabular}} &
  \textbf{\begin{tabular}[c]{@{}c@{}}Fully Controllable \\ Pointer Dereferencing\end{tabular}} \\ \midrule
Block     & \tickYes & 2 & \tickYes & \tickYes \\
Byte      & \tickYes & 2 & \tickYes & \tickYes \\
Event     & \tickYes & 1 & \tickYes & \tickYes \\
Mutex     & \tickNo  & 0 & \tickYes & \tickYes \\
Queue     & \tickYes & 2 & \tickYes & \tickYes \\
Semaphore & \tickYes & 1 & \tickYes & \tickYes \\
Thread    & \tickYes & 7 & \tickYes & \tickYes \\
Timer     & \tickYes & 3 & \tickYes & \tickYes \\ \bottomrule
\end{tabular}
\label{tab:type-for-malicious-and-accomplice-kernel-object}
\vspace{-3mm}
\end{table}

\paragraph{Performance of Symbolic Execution Engine (RQ4)}\label{rq4}
To evaluate the performance of our symbolic execution engine in identifying system call pairs, we recorded the performance metrics during the analysis of each system call. As shown in \autoref{tab:system-call-pairs-for-constructing-forged-kernel-objects}, most system calls were analyzed within one minute, thanks to our under-constrained approach which significantly reduces the exploration of unnecessary branches. 
However, \texttt{mutex\_put} required the longest time of almost 9 hours to explore 289 thousand paths. Our manual analysis of this system call revealed it involves numerous branches and loop operations, including checks on various mutex attributes and system states. We also observed that one system call (i.e., \texttt{thread\_priority\_change}) did not exit normally under our symbolic execution engine due to issues with constraint solving. We found this issue is caused by an under-constrained symbolized index of an array while it should be constrained in the kernel initialization stage in normal execution, which is ignored by our under-constrained symbolic execution. We will improve it in our future work.




\begin{table}[!t]
    \centering
    \scriptsize
    \aboverulesep=2pt
    \belowrulesep=2pt
    \setlength{\tabcolsep}{4pt}
    \setlength{\aboverulesep}{2pt}
    \caption{KOM Attacks on Different Platforms. \textit{*MPU} indicates disable MPU. \textit{*Read} indicates arbitrary memory read. \textit{*Write} indicates arbitrary memory write.}
\label{fig:KOM-attack-on-different-platforms}
\begin{tabular}{@{}cccccc@{}}
\toprule
\multicolumn{6}{c}{\textbf{KOM Attack on Different Platforms}}   \\ \midrule
\multicolumn{3}{c}{\textbf{Hardware Platform}}            & \multicolumn{3}{c}{\textbf{Attacks}}\\
\textbf{Board} & \textbf{Vendor} & \textbf{Core} & \textbf{*MPU} & \textbf{*Read} & \textbf{*Write} \\
NUCLEO-U575ZI-Q   & STM             & cortex-m33 & \tickYes & \tickYes & \tickYes \\
b-l475e-iot01a    & STM             & cortex-m4  & \tickYes & \tickYes & \tickYes \\
olimex-stm32-h405 & Olimex          & cortex-m4  & \tickYes & \tickYes & \tickYes \\
netduinoplus2     & Wilderness Labs & cortex-m4  & \tickYes & \tickYes & \tickYes \\ \bottomrule
\end{tabular}
\vspace{-3mm}
\end{table}

\paragraph{Implications of Attacks (RQ5)}\label{rq5}
To further demonstrate the impact of our attack, we launch KOM attacks on various platforms. As shown in \autoref{fig:KOM-attack-on-different-platforms}, we conducted the KOM attacks on these platforms, including disabling the MPU and achieving arbitrary memory read or write. This indicates that KOM attacks remain effective across different hardware platforms, further highlighting their broad impact. We provide a detailed Proof of Concept (PoC) for disabling MPU for interested readers in Appendix \S\ref{subsec:POC}. 


\section{Discussion}
\label{sec:Discussion}

\label{sub:mitigation}
\paragraph{Mitigation}
Given that all RTOSs expose kernel object pointers to threads, it's crucial to enhance kernel object validation. Following our discussion with AWS, FreeRTOS (v10.6.0) has implemented a table within the kernel memory to catalog information on kernel objects, including the kernel object pointers and their types. This enhancement aids in validating kernel object pointer parameters. For ThreadX, a similar strategy can be implemented to enhance the kernel object pointer validation.
\looseness=-1


In addition, most RTOSs expose all system calls, which expand the attack surface. They can mitigate this issue by reducing the system calls available to the threads. On the one hand, similar to Linux's seccomp\cite{DBLP:conf/uss/GhavamniaPMP20}, a set of filtering rules can be established to restrict access based on the importance and risk level of each system call. On the other hand, since developers have access to the full source code during the RTOS-based application development stage, they can choose to remove unnecessary system calls during the compilation phase.\looseness=-1

\paragraph{Noisy Memory Operations}
During the execution of the KOM attack, noisy memory operations may inadvertently corrupt sensitive fields within existing kernel objects. To address the issue, attackers can request two adjacent kernel objects (malicious and accomplice), which is likely to occur when there is sufficient allocation space with the default allocation algorithm in ThreadX. If not, the attacker can either retry the KOM attack or analyze the memory layout of kernel objects and the system call invocation patterns to minimize the noise.

\paragraph{Generality of the KOM attack}
Although the KOM attack has primarily targeted ThreadX, the methodology behind our attack is generalizable and can be applied to uncover other vulnerabilities in various RTOSs. For example, under-constrained symbolic execution can be utilized to explore system calls in other RTOSs, such as FreeRTOS, to identify vulnerabilities related to privilege separation, memory access control, and parameter sanitization. Additionally, we plan to explore the feasibility of conducting similar attacks on other RTOSs and general-purpose operating systems in future work.

\paragraph{Lessons Learnt}
The most important lesson from our study is that security protections require a comprehensive evaluation. First, it is crucial to assess the gap between the implementation of security protections and their actual effectiveness, particularly when these protections differ from those in full-fledged operating systems like Linux and Windows. While RTOSs strive to balance performance and security, the constraints of limited hardware resources and the emphasis on performance often result in customized security measures that can introduce unforeseen vulnerabilities. The KOM attack serves as a cautionary example for RTOS developers, underscoring the importance of a balanced approach to security and performance to prevent similar vulnerabilities.

Second, it is essential to consider the interdependencies among security protections. While RTOSs may implement multiple security measures, they often overlook the interplay between these protections, as discussed in \S\ref{lab:attack-surface-analysis}. The absence of any one security measure can weaken the effectiveness of others and undermine the security of the entire system.

Third, addressing security flaws in RTOSs requires more focused research and the development of automatic vulnerability detection techniques. Unlike full-fledged operating systems or Trusted Execution Environments (TEEs)~\cite{DBLP:conf/ccs/WangZHZGBLGC23,DBLP:conf/uss/CloostersRD20}, there has been limited effort in the security community to develop such detection tools for RTOSs. The unique software architectures and diverse hardware dependencies of RTOSs often render existing detection tools ineffective. Our work takes an initial step toward filling this gap by introducing an automated technique for identifying vulnerabilities. 
\looseness=-1

\section{Related work}

\paragraph{Pointer Sanitization} Over the past decade, parameter sanitization for pointers has been a focal point of concern within the security community, encompassing both general-purpose operating systems~\cite{DBLP:conf/apsys/ChenMWZZK11} and TEEs~\cite{optee-advisories,DBLP:journals/csur/PintoS19,DBLP:conf/ccs/BulckOMAGP19,DBLP:conf/ccs/WangZHZGBLGC23,DBLP:conf/uss/LeeJJKCCKPK17,DBLP:conf/uss/CloostersRD20,DBLP:conf/eurosp/KhanZKXBT23}. 
These attacks often exploit missing or improper pointer range checks in pointer sanitization, similar to the absence of pointer parameter sanitization in RTOSs, as observed in \S\ref{lab:attack-surface-analysis}. 
However, it is important to note that KOM attacks differ from these attacks. In our context, ThreadX leaks a kernel object pointer to a user thread, performs the pointer range check when the kernel object pointer is used in a system call, and performs semantic validation of object ID and some fields in the kernel object structure. ThreadX’s semantic validation and pointer range check of pointers would defeat these attacks. \looseness=-1

\paragraph{RTOS Security Protection} In addition to the inherent security protections outlined in \autoref{tab:protections-rtos}, recent research has been dedicated to developing supplementary, sophisticated security protections specifically for embedded applications on RTOSs~\cite{DBLP:conf/ccs/Tan023,DBLP:conf/uss/DuSD0WC22,DBLP:conf/sp/KhanXT23}. Furthermore, certain security protections~\cite{DBLP:conf/esorics/ShaoLLYWF22,DBLP:conf/uss/ZhouDSMCW20,DBLP:conf/ndss/AlmakhdhubCBP20,DBLP:conf/sp/ClementsASSKBP17} originally designed for bare-metal environments, can be adapted for use in RTOSs. However, these protections primarily focus on maintaining control flow integrity and often neglect the potential for data-only attacks~\cite{DBLP:conf/uss/YeLZ023,DBLP:conf/uss/JohannesmeyerSB24}. The parameter sanitization vulnerabilities identified in this paper can be exploited to conduct data-only attacks, enabling attackers to manipulate system call parameters and escalate privileges. Such exploits could compromise the security of the entire RTOS-based embedded system.\looseness=-1

\section{Conclusion}
\label{sec:Conclusion}

We discovered that a performance optimization in ThreadX introduces a vulnerability that can be exploited to bypass parameter sanitization. Our novel Kernel Object Masquerading attack can potentially lead to unauthorized data manipulation, privilege escalation, or even complete system compromise. To identify and evaluate KOM attacks, we developed an automated method based on under-constrained symbolic execution. Extensive experiments were conducted to validate the feasibility of KOM attacks on ThreadX-powered platforms.

\section*{Acknowledgment}
This research was supported in part by National Natural Science Foundation of China Grant Nos. 62232004 and 92467205 and 62441207, by US National Science Foundation (NSF) Awards 1931871 and 2325451, Jiangsu Provincial Key Laboratory of Network and Information Security Grant No. BM2003201, Key Laboratory of Computer Network and Information Integration of Ministry of Education of China Grant No. 93K-9, Natural Science Research Key Project of Universities by the Education Department of Anhui Province Grant No. 2024AH050146, and Collaborative Innovation Center of Novel Software Technology and Industrialization. Any opinions, findings, conclusions, and recommendations in this paper are those of the authors and do not necessarily reflect the views of the funding agencies.

\section*{Ethical Consideration}
We prioritize ethical considerations in our work. Firstly, all our experiments were conducted in controlled environments using our own devices. Secondly, we maintained strict confidentiality, ensuring that no information was shared with unauthorized individuals or organizations. Lastly, we responsibly disclosed the identified vulnerabilities, as detailed in \S\ref{para:vulnerabilities-in-mainstream-RTOSs}, to the respective vendors of FreeRTOS, LiteOS-M, and Mbed OS. For example, we reported the vulnerabilities of FreeRTOS to Amazon AWS and submitted a list of all system calls that could be exploited. These vulnerabilities garnered significant attention from AWS, leading to an invitation for an online meeting to collaboratively address the issue. Furthermore, AWS publicly acknowledged and thanked us on the official FreeRTOS GitHub repository.
For the KOM attack, we responsibly disclosed our findings to Microsoft. They confirmed the vulnerability and acknowledged us by posting our names on their website. Subsequently, they asked us to participate in the Microsoft Researcher Recognition Program, which would display our names on the MSRC websites.\looseness=-1

\section*{Open Science}
In compliance with the open science policy, we commit to openly sharing all research artifacts associated with our study. Given our research’s focus on security vulnerabilities and attack methodologies, we recognize the importance of transparency and accessibility in fostering a collaborative and robust security community.
To support replication and further investigation by peers, we make all artifacts related to our findings on RTOSs available~\cite{kom-experiment}. These artifacts include the under-constrained symbolic execution source code for the KOM attack, the PoC, and other relevant experimental data.

\bibliographystyle{IEEEtranS}
\bibliography{IEEEabrv,reference}
\appendix
\section{Appendix}

\subsection{Vulnerable Pointer Parameter Sanitization in FreeRTOS}
\label{appendix-a:weakpointercheck}

We discovered that FreeRTOS employs system calls with weak pointer parameter sanitization, leading to vulnerabilities that can be exploited to achieve privilege escalation and arbitrary memory access. We categorize these vulnerabilities into two main types: direct out-of-bound pointer dereference ($V1$) and indirect out-of-bound pointer dereference ($V2$). The first type ($V1$) occurs when an attacker passes a malicious pointer parameter that points to privileged memory from a compromised unprivileged thread. The second type ($V2$) occurs when the attacker passes a parameter that can indirectly manipulate a pointer to access privileged memory.\looseness=-1

\begin{figure}[ht]
\centering
\includegraphics[width=\linewidth]{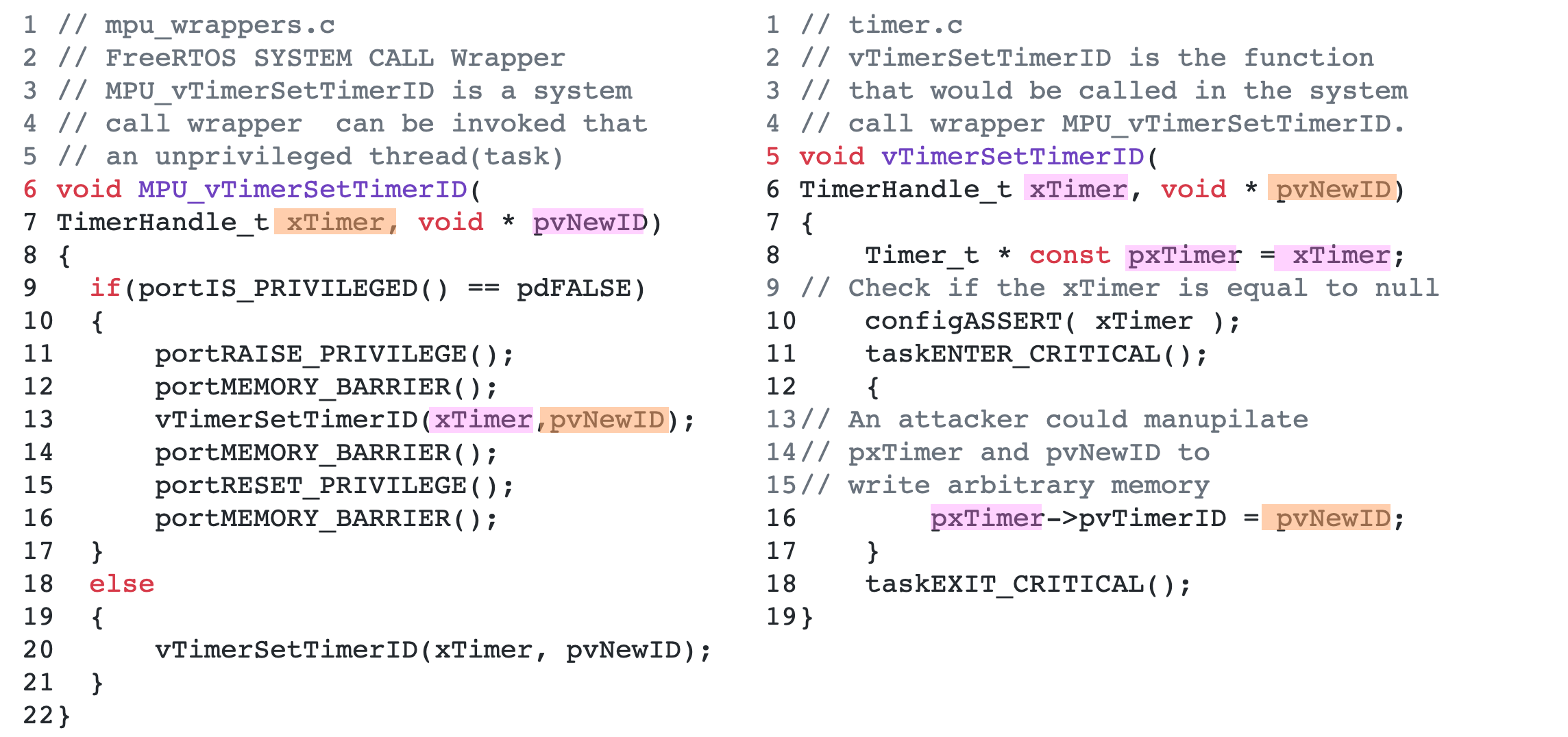}

\caption{The Sample Code of Direct Out-of-bound Pointer Dereference}
\label{fig:freertos-v1-code-example}

\end{figure}

\begin{figure}[ht]
\centering
\includegraphics[width=\linewidth]{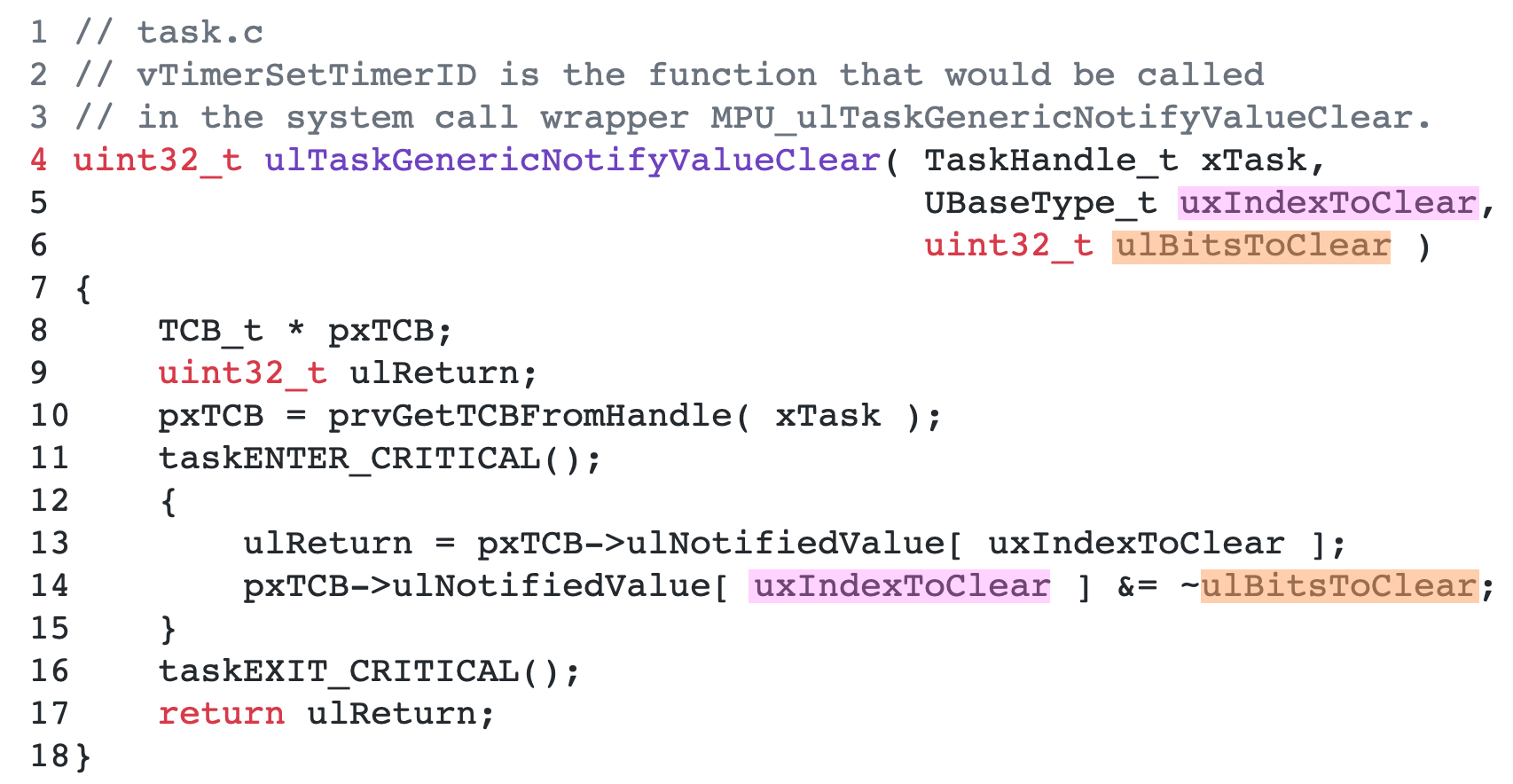}
\caption{The Sample Code of Indirect Out-of-bound Pointer Dereference}
\label{fig:freertos-v2-code-example}
\end{figure}

For example, \autoref{fig:freertos-v1-code-example} illustrates $V1$, which enables an attacker to access arbitrary memory. The FreeRTOS kernel escalates the privilege if the system call \texttt{MPU\_vTimerSetTimerID(.)} is invoked from an unprivileged thread and then call \texttt{vTimerSetTimerID(.)}. An attacker can manipulate the parameters (i.e., \texttt{xTimer} and \texttt{pvNewID}) to write arbitrary memory, as \texttt{vTimerSetTimerID(.)} dereferences the pointer provided by \texttt{xTimer} and updates it with the value of \texttt{pvNewID}.\looseness=-1

\autoref{fig:freertos-v2-code-example} presents an example of $V2$. In contrast to $V1$, here an attacker can alter the memory location pointed to by an array's index parameter (i.e., \texttt{uxIndexToClear}), and subsequently modify the contents at the targeted memory address using the \texttt{ulBitsToClear} parameter.\looseness=-1

\subsection{Proof of Concept}
\label{subsec:POC}
To demonstrate the effectiveness of our attack, we show the KOM attack can turn off the MPU by exploiting the KOM vulnerability. We choose an official sample project provided by STM, named \texttt{Tx\_MPU}, as the target for our POC. This project can serve as a template for the development of numerous applications for embedded systems. We deploy this project on an STM32U575 development board and enable the MPU protection. Before detailing our specific attack, we assume that an attacker has compromised a thread and can invoke any system calls. 
Our exploitation allows the malicious unprivileged thread to write the MPU control register and consequently disable the MPU by leveraging the following six system calls, i.e., \texttt{module\_object\_allocate}, \texttt{timer\_create}, \texttt{byte\_allocate}, \texttt{thread\_create}, \texttt{thread\_time\_slice\_change},  \texttt{thread\_reset}. \texttt{thread\_time\_slice\_change} assigns the second parameter to a data member variable (\texttt{thread\_new\_time\_slice}) of the specified thread object. The system call \texttt{thread\_reset} can dereference a pointer (i.e., \texttt{thread\_stack\_start}) and write a fixed value. Before the dereference, \texttt{thread\_reset} would validate \texttt{thread\_stack\_size} and \texttt{thread\_state} as in semantic validation. Our PoC leverages KOM attack to overwrite the \texttt{thread\_stack\_start} of a thread object with the address of the MPU control register and then calls \texttt{thread\_reset} to disable the MPU.  
To be more specific, there are a total of three steps in this PoC attack:

\begin{figure}[ht]
\centering
\includegraphics[width=\linewidth]{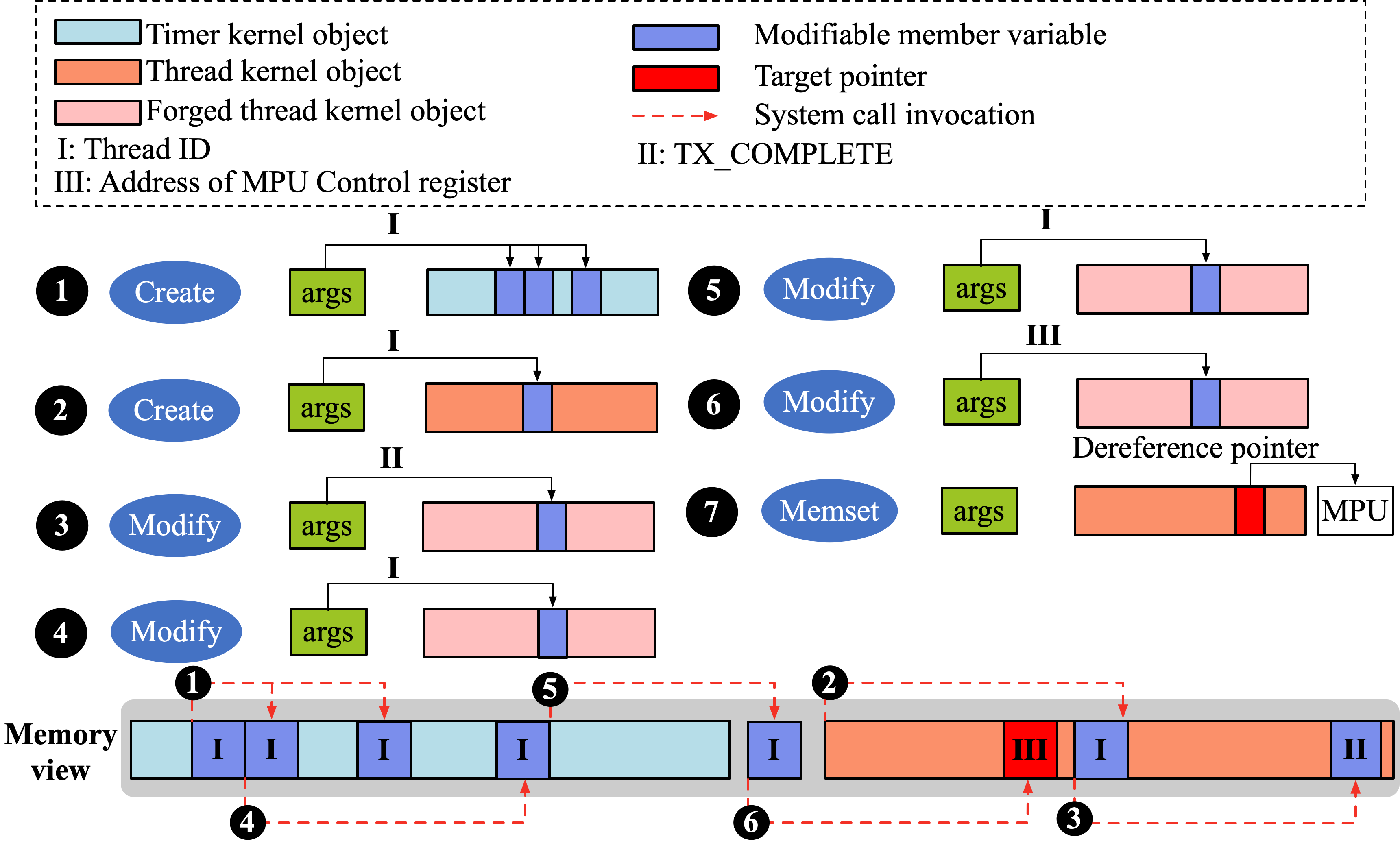}
\caption{Compromise MPU Flow}
\label{fig:compromise-mpu-flow}

\end{figure}

\paragraph{Step I - Creation of Malicious Kernel Objects and Selection of Accomplice Kernel Objects} 
We use \texttt{module\_object\_allocate} and \texttt{timer\_create} to create a timer object named \texttt{MaliciousTimer} (i.e., malicious kernel object), and use \texttt{module\_object\_allocate} and \texttt{thread\_create} to create a thread object named \texttt{MaliciousThread} (i.e., accomplice kernel object). 
When creating \texttt{MaliciousTimer} via calling \texttt{timer\_create}, we specifically pass the value \texttt{0x54485244}, which is the type ID of a thread object (i.e., Thread ID), as the 4th, 5th and 6th parameters of \texttt{timer\_create}. These three parameters are assigned to three data member variables of \texttt{MaliciousTimer} respectively (\ding{182} in \autoref{fig:compromise-mpu-flow}). Similarly, we pass a value \texttt{DEFAULT\_STACK\_SIZE} and a value \texttt{0x54485244} as the 6th and last second parameter to \texttt{thread\_create}. The first value is used for \texttt{thread\_stack\_size} while the second value is a valid Thread ID for constructing a forged thread object later (\ding{183} in \autoref{fig:compromise-mpu-flow}).

\begin{figure}[ht]
\vspace{-3mm}
\centering
\includegraphics[width=\linewidth]{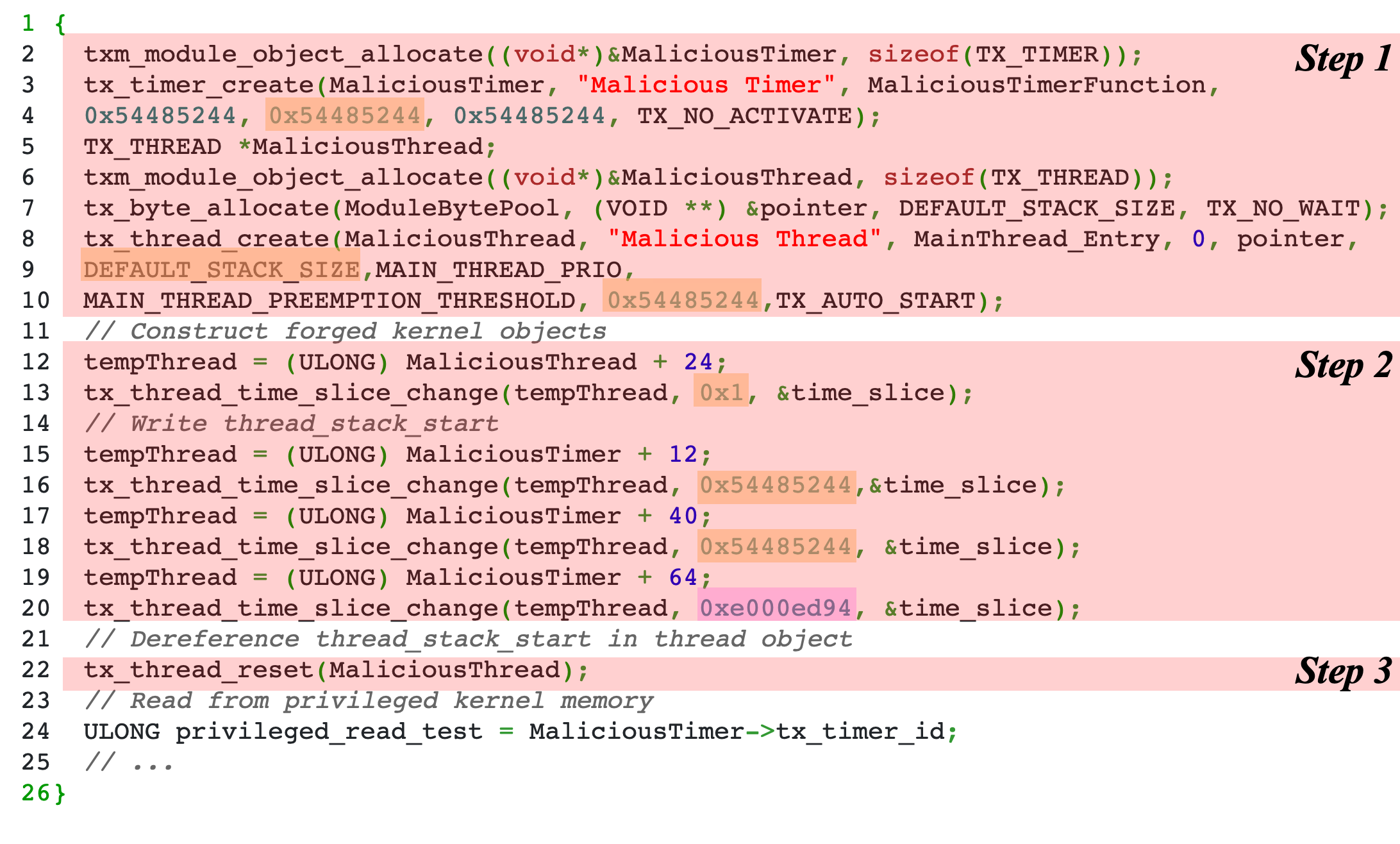}
 
\caption{Exploitation Code of Compromising MPU}
\label{fig:compromise-mpu-code}
\vspace{-3mm}
\end{figure}

\paragraph{Step II - Manipulation of Accomplice Kernel Object's pointer}
To manipulate the pointer of the accomplice kernel object. We first call \texttt{thread\_time\_slice\_change} to construct a forged thread object to modify the \texttt{thread\_state} of the accomplice kernel object, which is one of the characteristic fields of \texttt{thread\_reset} before dereferencing the pointer (\ding{184} in \autoref{fig:compromise-mpu-flow}).
Then, by calling \texttt{thread\_time\_slice\_change} two times, we can construct two forged thread objects as shown in (\ding{185}, \ding{186} in \autoref{fig:compromise-mpu-flow}) based on the \texttt{MaliciousTimer}.
After that, we call \texttt{thread\_time\_slice\_change}, passing 0xe000ed94 (the desired malicious pointer value) as the second parameter.
It modifies the modifiable field \texttt{thread\_new\_time\_slice} of the corresponding kernel object with 0xe000ed94. This data member variable overlaps the pointer \texttt{thread\_stack\_start} of \texttt{MaliciousThread}. The value 0xe000ed94 is actually the address of the MPU control register.
\looseness=-1

\paragraph{Step III - Disabling MPU  via pointer Dereference}
We call \texttt{thread\_reset} with \texttt{MaliciousThread} as the sole parameter to dereference the pointer \texttt{thread\_stack\_start} of \texttt{MaliciousThread} and overwrite the MPU control register. To check whether MPU has truly been disabled, we can directly read a member variable that is inaccessible for a thread, i.e., \texttt{MaliciousTimer->tx\_timer\_id}. 


\end{document}